\documentclass[twocolumn]{aastex62}

\newcommand {\kms}{$\rm{km\ s^{-1}}$}
\newcommand{\ergcms}{$\rm ~erg~cm^{-2}~s^{-1}$}
\newcommand{\ccm} {{\rm cm}^{-3}}
\newcommand{\gccm} {{\rm g \ cm}^{-3}}

\received{September 16, 2019}
\revised{October 18, 2019}
\accepted{October 18, 2019}

\shorttitle{HST observations of SN 1987A}
\shortauthors{Larsson et al.}

\begin{document}

\title{The matter beyond the ring: the recent evolution of SN 1987A observed by the Hubble Space Telescope}

\correspondingauthor{J. Larsson}
\email{josla@kth.se}

\author[0000-0003-0065-2933]{J. Larsson}
\affil{Department of Physics, KTH Royal Institute of Technology, The Oskar Klein Centre, AlbaNova, SE-106 91 Stockholm, Sweden}

\author[0000-0001-8532-3594]{C. Fransson}
\affiliation{Department of Astronomy, Stockholm University, The Oskar Klein Centre, AlbaNova, SE-106 91 Stockholm, Sweden}

\author[0000-0002-0427-5592]{D. Alp}
\affil{Department of Physics, KTH Royal Institute of Technology, The Oskar Klein Centre, AlbaNova, SE-106 91 Stockholm, Sweden}

\author{P. Challis}
\affil{Harvard-Smithsonian Center for Astrophysics, 60 Garden Street, MS-19, Cambridge, MA 02138, USA}

\author[0000-0002-9117-7244]{R. A. Chevalier}
\affiliation{Department of Astronomy, University of Virginia, P.O. Box 400325, Charlottesville, VA 22904-4325, USA}

\author[0000-0002-1002-3674]{K. France}
\affil{Laboratory for Atmospheric and Space Physics, University of Colorado, 392 UCB, Boulder, CO 80309, USA}

\author{R. P. Kirshner}
\affil{Harvard-Smithsonian Center for Astrophysics, 60 Garden Street, MS-19, Cambridge, MA 02138, USA}
\affil{Gordon and Betty Moore Foundation, 1661 Page Mill Road, Palo Alto, CA 94304, USA}

\author[0000-0002-7491-7052]{S. Lawrence}
\affil{Department of Physics and Astronomy, Hofstra University, Hempstead, NY 11549, USA}

\author[0000-0002-4413-7733]{B. Leibundgut}
\affiliation{European Southern Observatory, Karl-Schwarzschild-Strasse 2, D-85748 Garching, Germany}

\author[0000-0002-3664-8082]{P. Lundqvist}
\affiliation{Department of Astronomy, Stockholm University, The Oskar Klein Centre, AlbaNova, SE-106 91 Stockholm, Sweden}

\author{S.  Mattila}
\affiliation{Tuorla observatory, Department of Physics and Astronomy, University of Turku, Väisäläntie 20, FI-21500 Piikkiö, Finland}

\author[0000-0002-8855-8679]{K.  Migotto}
\affiliation{Department of Astronomy, Stockholm University, The Oskar Klein Centre, AlbaNova, SE-106 91 Stockholm, Sweden}

\author[0000-0003-1546-6615]{J. Sollerman}
\affiliation{Department of Astronomy, Stockholm University, The Oskar Klein Centre, AlbaNova, SE-106 91 Stockholm, Sweden}

\author[0000-0003-1440-9897]{G. Sonneborn}
\affiliation{NASA/Goddard Space Flight Center, Observational Cosmology Lab, Code 665, Greenbelt, MD 20771}

\author[0000-0001-6815-4055]{J. Spyromilio}
\affiliation{European Southern Observatory, Karl-Schwarzschild-Strasse 2, D-85748 Garching, Germany}

\author{N. B. Suntzeff}
\affiliation{George P. and Cynthia Woods Mitchell Institute for Fundamental Physics \& Astronomy, Texas A\&M University, Department of Physics, 4242 TAMU, College Station, TX 77843}

\author[0000-0003-1349-6538]{J. C. Wheeler}
\affiliation{Department of Astronomy, University of Texas, Austin, TX 78712-0259, USA}

\begin{abstract}

The nearby SN 1987A offers a spatially resolved view of the evolution of a young supernova (SN) remnant. Here we precent recent {\it Hubble Space Telescope} imaging observations of SN 1987A, which we use to study the evolution of the ejecta, the circumstellar equatorial ring (ER) and the increasing emission from material outside the ER.  We find that the inner ejecta have been brightening at a gradually slower rate and that the western side has been brighter than the eastern side since $\sim 7000$~days. This is expected given that the X-rays from the ER are most likely powering the ejecta emission. At the same time the optical emission from the ER continues to fade linearly with time. The ER is expanding at $680\pm 50$~\kms, which reflects the typical velocity of transmitted shocks in the dense hotspots. A dozen spots and a rim of diffuse H$\alpha$ emission have appeared outside the ER since $9500$~days. The new spots are more than an order of magnitude fainter than the spots in the ER and also fade faster. We show that the spots and diffuse emission outside the ER may be explained by fast ejecta interacting with high-latitude material that extends from the ER toward the outer rings. Further observations of this emission will make it possible to determine the detailed geometry of the high-latitude material and provide insight into the formation of the rings and the mass-loss history of the progenitor. 

\end{abstract}

\keywords{Core-collapse supernovae (304) -- Supernova remnants (1667) -- Circumstellar matter (241) -- Shocks (2086) -- Ejecta (453) }

\section{Introduction}
\label{sec:intro}

SN~1987A in the Large Magellanic Cloud (LMC) was discovered on 23 February 1987. It was a naked eye object and the closest SN since Kepler's SN in 1604 (see \citealt{McCray1993}  and \citealt{McCray2016} for reviews). The {\it Hubble Space Telescope (HST)} was launched a few years after this historic event, making it possible to follow the evolution from SN to SN remnant (SNR) with excellent spatial resolution. The {\it HST} images offer a resolved view of the ejecta, which are expanding into a circumstellar  triple-ring nebula. 

The three rings of the nebula are nearly circular with inclinations in the range $\sim 38-45^{\circ}$  \citep{Crotts2000,Tziamtzis2011}. The inner, equatorial ring (ER) has a radius of $\sim 0.6$~lt-yr, while the two outer rings (ORs) are $\sim 2.2$ times bigger. The expansion velocities of the rings suggest that they all formed around the same time,  $\sim 20,000$ years before the explosion \citep{Crotts2000}. Many scenarios have been proposed to explain the origin of the rings, including interaction of winds from different evolutionary stages of the progenitor star \citep{Wang1992, Blondin1993,Martin1995}, a binary merger \citep{Morris2007,Morris2009}, bipolar outflows/jets \citep{Akashi2015} and the remains of a protostellar disc \citep{McCray1994}. Photoionization of the wind medium and/or photoevaporation of the ER by the radiation from the progenitor may also have been important in shaping the nebula \citep{Chevalier1995,Smith2013}. The latter was suggested as an explanation for the blue supergiant (BSG) SBW1, which has a ring system that resembles SN~1987A \citep{Smith2013}. Similar ring systems have also been observed around the BSGs  Sher~25 \citep{Brandner1997} and HD~168625 \citep{Smith2007}. 

Understanding the formation of the rings is connected to understanding the progenitor.  In the case of SN~1987A, the progenitor was a B3 Ia BSG \citep{Walborn1987,West1987}. 
Mass estimates assuming a single-star progenitor are in the range $\sim 15-20~\rm{M}_{\odot}$ (\citealt{Woosley1988,Fransson2002,Smartt2009,Pumo2011,Orlando2015,Utrobin2019}), although \cite{Utrobin2019} notes that there is no model that satisfies all observational constraints. Binary merger models suggest a somewhat larger mass \citep{Menon2017}. The progenitor makes SN~1987A different from the majority of type II SNe, which originate from red supergiants \citep{Smartt2009}. Specifically, the relatively compact nature of the BSG progenitor resulted in a $^{56}$Ni dominated peak in the light curve rather than a plateau. A number of other SN~1987A-like events have been observed and the rate has been estimated to 1-3$\%$ of the core collapse rate \citep{Pastorello2012,Taddia2016}. There is observational evidence that these SNe preferentially occur in low-metallicity environments \citep{Pastorello2012,Taddia2013}, which may explain explosions in the BSG stage \citep{Arnett1989}. Alternatively, the BSG progenitor of SN~1987A may  be explained by a binary merger (e.g., \citealt{Podsiadlowski1990,Podsiadlowski1992,Menon2017}). Further insights into the explosion would be obtained from the properties of the compact object, but this has not been detected subsequent to the original neutrino emission \citep{Alp2018}.

The ring system was initially ionized by the SN flash and then faded with time as the gas recombined and cooled \citep{Fransson1989,Lundqvist1996,Mattila2010}. After $\sim 3000$ days the ejecta reached the ER, resulting in the appearance of optical ``hotspots" \citep{Sonneborn1998,Lawrence2000}. Prior to this, radio and X-ray observations had revealed emission originating from the interaction between the ejecta and a low-density ($\sim10^2\ \rm{cm^{-3}}$) H~II region inside the ER  \citep{Staveley1992,Chevalier1995,Hasinger1996}. Subsequent multi-wavelength monitoring showed that the interaction with the ER gave rise to rapidly increasing emission across the entire electromagnetic spectrum for many years (e.g., \citealt{Ng2008,Groningsson2008b,Maggi2012}). The optical emission finally peaked after 8000~days and has since been fading, signaling that the blast wave has left the ER and that the clumps are being destroyed by the shocks \citep{Fransson2015}. This evolution has been accompanied by the appearance of faint spots outside the ER in the {\it HST} images after 9500~days. Observations at other wavelengths have also revealed transitions in the evolution of the ER, including fading in the IR \citep{Arendt2016}, a flattening of soft X-rays \citep{Frank2016} and a re-acceleration of the blast wave in radio \citep{Cendes2018}. All these observations show that the remnant of SN~1987A has entered a new phase. Hydrodynamic simulations of the ring collision are also broadly consistent with this picture  \citep{Orlando2015}.

The ejecta inside the ER are expanding freely, but are nevertheless being affected by the shock interaction in the sense that the X-ray emission from the ER serves as an important energy source. This started dominating over the $^{44}$Ti decay after $\sim 5000$ days, leading to a brightening of the optical emission from the ejecta \citep{Larsson2011}. Most of the X-rays are expected to be absorbed in the outer parts of the freely expanding ejecta, which leads to an edge-brightened morphology \citep{Larsson2013,Fransson2013}. By contrast, the $^{44}$Ti decay still dominates the energy input to the innermost ejecta, as evidenced by the constant flux of the 1.644~$\mu$m [Fe~I]+[Si~II] line \citep{Larsson2016}. The fact that molecules and dust are observed in the inner ejecta also shows that these regions are not significantly affected by the X-rays \citep{Matsuura2015,Fransson2016,Abellan2017,Larsson2019,Cigan2019}. The 3D emissivities of a number of atomic and molecular emission lines from the ejecta have been determined using a combination of imaging and spectra, revealing a highly asymmetric distribution \citep{Kjaer2010,Larsson2013,Larsson2016,Abellan2017,Larsson2019}. Several different early observations also provided strong evidence for an asymmetric explosion, including  polarimetry \citep{Schwarz1987,Jeffery1991}, the ``mystery spots" \citep{Nisenson1987,Meikle1987,Nisenson1999} and  the ``Bochum event"  \citep{Hanuschik1990}. 

In this paper we present the recent evolution of SN~1987A as observed by {\it HST}, focussing on the time period between 10,300 and 11,500 days (years 2015 to 2018). This extends  the decaying part of the light curve of the ER compared to  the data in \cite{Fransson2015} and reveals new details about the structure of the circumstellar medium outside the ER. After describing the observations in Section~\ref{sec:obs}, we give an overview of the optical emission components in Section~\ref{sec:optem}, show images in Section~\ref{sec:images}, and present light curves for the ER and ejecta in Section~\ref{sec:lightcurves}. We then analyze the expansion of the ER and the properties of the new spots outside the ER in Sections~\ref{sec:expansion} and \ref{sec:spots}, respectively. We finally discuss the results in Section~\ref{sec:disc} and summarize our conclusions in Section~\ref{sec:conclusions}. Throughout this paper we assume a distance to the LMC of 49.6~kpc \citep{Pietrzynski2019}. All uncertainties are quoted at 1$\sigma$.

\section{Observations and data reduction}
\label{sec:obs}

\begin{deluxetable}{cccccc}[t]
\tablecaption{HST observations in the F502N filter  \label{tab:f502nobs}}
\tablecolumns{4}
\tablenum{1}
\tablewidth{0pt}
\tablehead{
\colhead{Date} & 
\colhead{Epoch\tablenotemark{a}} & 
\colhead{Instrument} &
\colhead{Exposure time} \\
\colhead{(YYYY-mm-dd)} &
\colhead{(d)} &
\colhead{} &
\colhead{(s)} 
}
\startdata
1994-02-03 & 2537 & WFPC2  & 5600  \\
1996-02-06 & 3270 & WFPC2  & 7800  \\
1997-07-12 & 3792 & WFPC2  & 8200  \\
2000-06-16 & 4862 & WFPC2  & 3600  \\
2000-11-14 & 5013 & WFPC2  & 5600  \\
2001-12-07 & 5401 & WFPC2  & 4800  \\
2003-01-05 & 5795 & ACS  & 7000  \\
2003-11-28 & 6122 & ACS  & 4000  \\
2004-12-15 & 6505 & ACS  & 3600  \\
2005-11-18 & 6843 & ACS  & 4240  \\
2006-12-08 & 7228 & ACS  & 2600  \\
2009-12-12 & 8328 & WFC3  &  3100 \\
2014-06-20 & 9979 & WFC3  &  5880 \\
2016-06-08 & 10,698 & WFC3 & 2400  \\
2017-08-03 & 11,119 & WFC3  & 3300  \\
2018-07-10& 11,460  & WFC3 & 5760  \\
\enddata
\tablenotetext{a}{Days since explosion on 1987-02-23.}
\end{deluxetable}

\begin{deluxetable}{cccccc}[t]
\tablecaption{HST/WFC3 observations since 2015 \label{tab:w3obs}}
\tablecolumns{4}
\tablenum{2}
\tablewidth{0pt}
\tablehead{
\colhead{Date} & 
\colhead{Epoch\tablenotemark{a}} & 
\colhead{Filter} &
\colhead{Exposure time} \\
\colhead{(YYYY-mm-dd)} &
\colhead{(d)} &
\colhead{} &
\colhead{(s)} 
}
\startdata
2015-05-24 & 10,317 & F438W  & 1200  \\
		  & 		 & F625W  & 1200  \\
2016-06-08 & 10,698 & F438W  & 600  \\
                   & 		 & F625W  & 600  \\
                   &		& F645N  &  2400 \\
                   & 		 & F657N  & 2400  \\
2017-08-03 & 11,119 & F438W  & 1400  \\
		  & 		 & F625W  & 1200  \\
		  & 		 & F657N  & 2800  \\
2018-07-08 &  11,458 & F438W  & 1200  \\
		  & 		 & F625W  & 1200  \\
2018-07-10 & 11,460 & F657N  & 2880  \\
\enddata
\tablecomments{The observations in the F502N filter are listed in Table \ref{tab:f502nobs}.}
\tablenotetext{a}{Days since explosion on 1987-02-23.}
\end{deluxetable}

\begin{deluxetable}{ccccccc}[t]
\tablecaption{Filter flux ratios  \label{tab:corrfactors}}
\tablecolumns{6}
\tablenum{3}
\tablewidth{0pt}
\tablehead{
\colhead{Instrument} & 
\multicolumn2c{B band\tablenotemark{a}} & 
\multicolumn2c{R band\tablenotemark{b}} &
\colhead{F502N} \\
\colhead{} &
\colhead{ER} &
\colhead{ejecta} &
\colhead{ER} &
\colhead{ejecta} &
\colhead{ER} 
}
\startdata
ACS & 1.38 & 1.26 & 1.06 & 1.05 &  0.87    \\
WFC3 & 1.00 & 1.07 & 1.13 & 1.13 &  0.82 \\
\enddata
\tablecomments{Flux ratios relative to the corresponding filters used with WFPC2. All light curves have been corrected by these factors and hence normalized to the bandpass of the WFPC2 filters.}
\tablenotetext{a}{WFPC2/F439W, ACS/F435W, WFC3/ F438W}
\tablenotetext{b}{WFPC2/F675W, ACS/F625W, WFC3/F625W}
\end{deluxetable}

\begin{deluxetable*}{cccccc}[t]
\tablecaption{Strongest emission lines from the ER and ejecta affecting the different filters \label{tab:lines}}
\tablecolumns{3}
\tablenum{4}
\tablewidth{0pt}
\tablehead{
\colhead{Filter} & 
\colhead{Main lines from the ejecta\tablenotemark{a}} & 
\colhead{Main lines from the ER} &
}
\startdata
F438W  & Ca~I $\lambda 4228$, Fe~I lines, Mg~I] $\lambda 4571$ 	&	[S~II] $\lambda 4069$, H$\delta$, H$\gamma$, [Fe~II] lines \\
F502N  & Fe~I $\lambda 5007$ [$-2400,+2300$]~\kms  & 									*[O~III] $\lambda 5007$, He~I	 $\lambda 5016$\\
F625W   & Na~I  $\lambda \lambda 5890, 5896$, [O~I] $\lambda \lambda 6300, 6364$,   	& [N~II] $\lambda 5755$, He~I $\lambda 5876$, [O~I] $\lambda \lambda 6300, 6364$,\\
	     &  *H$\alpha$  & *H$\alpha$, [N~II] $\lambda \lambda 6548, 6583$	 \\
F645N   & *H$\alpha$ [$-7900, -2700$]~\kms & Continuum\\	     
F657N   & *H$\alpha$ [$-3500, +3600$]~\kms & *H$\alpha$, [N~II] $\lambda \lambda 6548, 6583$	\\
\enddata
\tablecomments{Lines that dominate the emission in a filter are indicated by *. }
\tablenotetext{a} {For each of the narrow filters we provide the velocity range of the strongest line included in the filter at the systematic velocity of SN~1987A (287~\kms, \citealt{Groningsson2008b}). The limits correspond to 10$\%$ of the peak transmission.}
\end{deluxetable*}

The details of the {\it HST} observations of SN~1987A analyzed in this paper are provided in Tables \ref{tab:f502nobs} and \ref{tab:w3obs}. Table~\ref{tab:f502nobs} contains all observations in the F502N  filter between 1994 and 2018, while Table~\ref{tab:w3obs} lists WFC3 observations since 2015, including F438W (B band), F625W (R band), F645N and F657N. In addition, we use all previous observations in the R band (WFPC2/F675W, ACS/F625W, WFC3/F625W) and  B band (WFPC2/F439W, ACS/F435W, WFC3/ F438W), as well as a WFC3/F657N image from 2011 and a WFC3/F645N image from 2014. These observations have been discussed in \cite{Larsson2011},  \cite{Fransson2015} and \cite{France2015}. 

All observations since the year 2000 are based on dithered exposures, where offsets involving half-integer pixels are used to improve the sampling. This leads to an improved spatial resolution when the images from different exposures are combined. We combined the images using DrizzlePac\footnote{\url{http://drizzlepac.stsci.edu}}, which also removes cosmic rays and applies distortion corrections \citep{Gonzaga2012}. All the images were drizzled to a pixel scale of $0\farcs{025}$ and aligned using the positions of stars located around SN~1987A. 

The comparison of fluxes from SN~1987A at different epochs is affected by the fact that the observations have been performed with three different instruments (WFPC2, ACS and WFC3), for which the relevant filters differ somewhat in terms of wavelength coverage and throughput. To quantify this effect we convolve the spectra of the ER and the ejecta with the response functions of the different instrument/filter combinations and compare the resulting fluxes. We use the spectra from \cite{Fransson2013} and \cite{Larsson2013}, and calculate the fluxes with STSDAS SYNPHOT.\footnote{\url{http://www.stsci.edu/institute/software_hardware/stsdas/synphot}} From the results we determine the  correction factors in Table~\ref{tab:corrfactors}, which are normalized with respect to the WFPC2 filters. All light curves presented below have been corrected by these factors. The largest corrections apply to the ACS B band, which is significantly wider than the corresponding B-band filters used with WFPC2 and WFC3. There is a small uncertainty in the correction factors due to the moderate time evolution of the spectrum of SN~1987A.  We investigated this using spectra from a selection of epochs relevant for the different instruments (obtained from \citealt{Groningsson2008b,Fransson2013}, and K.~Migotto et al., in preparation). This showed that that the spectral evolution affects the correction factors by $\la 1\%$, with the exception of the ACS B band, where the maximal difference between epochs was $4\%$. 

A small number of the observations are significantly affected by the degradation of the charge transfer efficiency (CTE), which means that trapping of charges during CCD readout results in lower fluxes being measured. In the case of our measurements this primarily affects the three WFPC2 observations between 7400--8100 days. This is the time after the ACS had failed, but before WFC3 was commissioned. Correction formulas for CTE losses exist for isolated point sources, but these are not appropriate for SN 1987A, which consists of a combination of diffuse emission and closely spaced point sources. Instead, we simply apply empirical corrections to the light curves, as described in Section~\ref{sec:lightcurves}. We note that the precise fluxes in these previous observations do not affect our conclusions.

\section{Optical emission from SN 1987A}
\label{sec:optem}

\begin{figure}[t]
\plotone{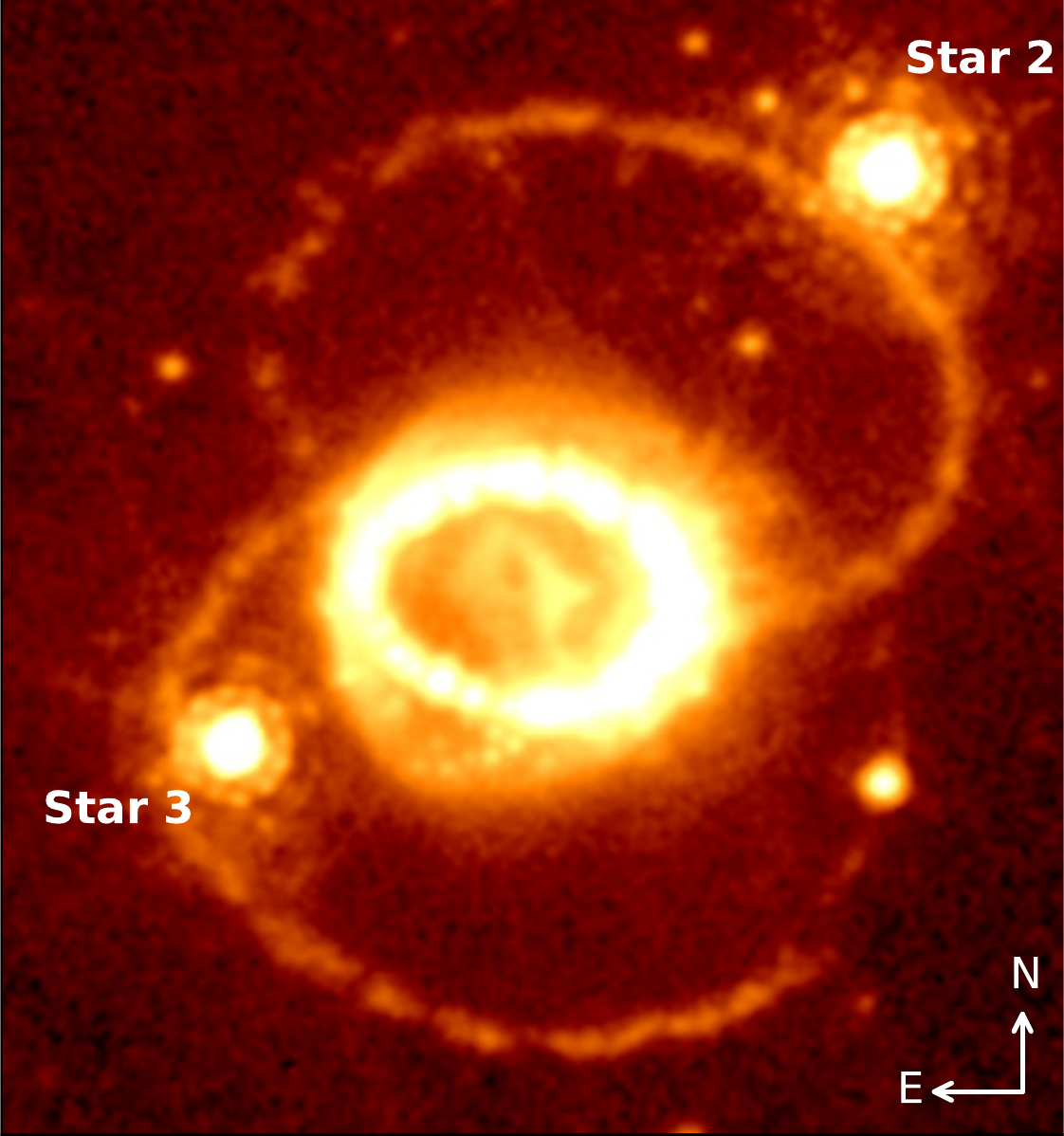} 
\caption{WFC3/F657N image from 11,500~days (2018) showing the triple-ring system around SN~1987A. The image scale is logarithmic in order to highlight the ORs and the faint emission outside the ER. The field of view (FOV) is $5\farcs{5} \times 5\farcs{9}$.\\\label{fig:3ring}}
\end{figure}
\begin{figure}[t] 
\plotone{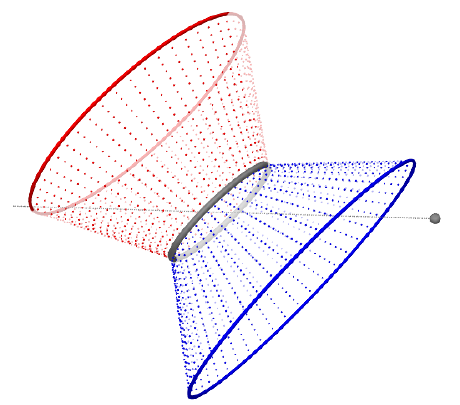} 
\caption{A 3D view of the ring system surrounding SN~1987A.  The ER is shown in grey, the NOR in red and the SOR in blue. The dashed lines represent hypothetical material connecting the ER and ORs. The grey sphere and line show the position of the observer and the line of sight, respectively. \\\label{fig:3ringmod}}
\end{figure}

The optical flux from SN~1987A is primarily due to emission lines (see e.g., \citealt{Pun2002,Michael2003,Larsson2013,Fransson2013}). In Table~\ref{tab:lines} we list the strongest lines from the ER and ejecta that contribute to the broad and narrow filters analyzed in this paper. We note that the R band and F657N are strongly dominated by H$\alpha$, and that the emission from the ER in the F502N filter is dominated by [O~III] $\lambda 5007$. The F645N filter includes blueshifted H$\alpha$ emission as well as a weak continuum due to H~I and He~I two-photon and free-bound emission from the ER. 

 In order to illustrate all the different emission sites in the remnant, we show the most recent image in the F657N filter in Figure~\ref{fig:3ring}. This shows the triple-ring nebula, where the ER is significantly brighter than the ORs due to the interaction with the ejecta.  A schematic 3D view of the ring system is provided in Figure~\ref{fig:3ringmod}.  The northern OR (NOR) is on the far side of the ER and the southern OR (SOR) on the near side with respect to the observer. There may also be material connecting the ER with the ORs as illustrated by the dashed lines in Figure~\ref{fig:3ringmod}. The details of the geometry of the ring system will be discussed in Section~\ref{sec:disc-outside}. The two bright stars located to the north-west and south-east of the SN are usually referred to as Star~2 and Star~3, respectively (\citealt{Walborn1987}, labeled in Figure~\ref{fig:3ring}). 

The bright asymmetric structure seen inside the ER in Figure~\ref{fig:3ring} is the freely expanding inner ejecta. The outer parts of the ejecta are interacting with a reverse shock (RS), which gives rise to diffuse emission that extends from the inner edge of the ER to higher latitudes above and below its plane. The RS is probably the main origin of the arc of emission seen just inside the western side of the ER, although scattered light from the ER is also expected to contribute. The ER  itself  is composed of individual hotspots, as seen in Figures~\ref{fig:allred} and \ref{fig:allblue} below. There are also a number of faint spots located outside the ER to the south-east. In this paper we will show that the spots and diffuse emission outside the ER likely originate in material that extends from the ER toward the ORs. 
 
The inner ejecta emit lines with typical FWHM $\sim 2500$~\kms. Spectra of the ER contain three emission components:  a very broad component extending to $\sim~10,000$~\kms\ from the RS, lines of intermediate width (FWHM $\sim~200-450$~\kms) from the shocked gas in the hotspots, and narrow lines (FWHM $\sim~10-30$~\kms) from the unshocked, photoionized gas \citep{Groningsson2008b,Fransson2013}. The latter have been fading since very early times and have almost disappeared at current epochs (K. Migotto et al. 2019, in preparation). The analysis presented below will focus on the evolution of the ejecta, the ER and the new emission outside the ER. The evolution of the ORs has been studied in \cite{Tziamtzis2011}.

\section{Images}
\label{sec:images}

\begin{figure*}[t]
\plotone{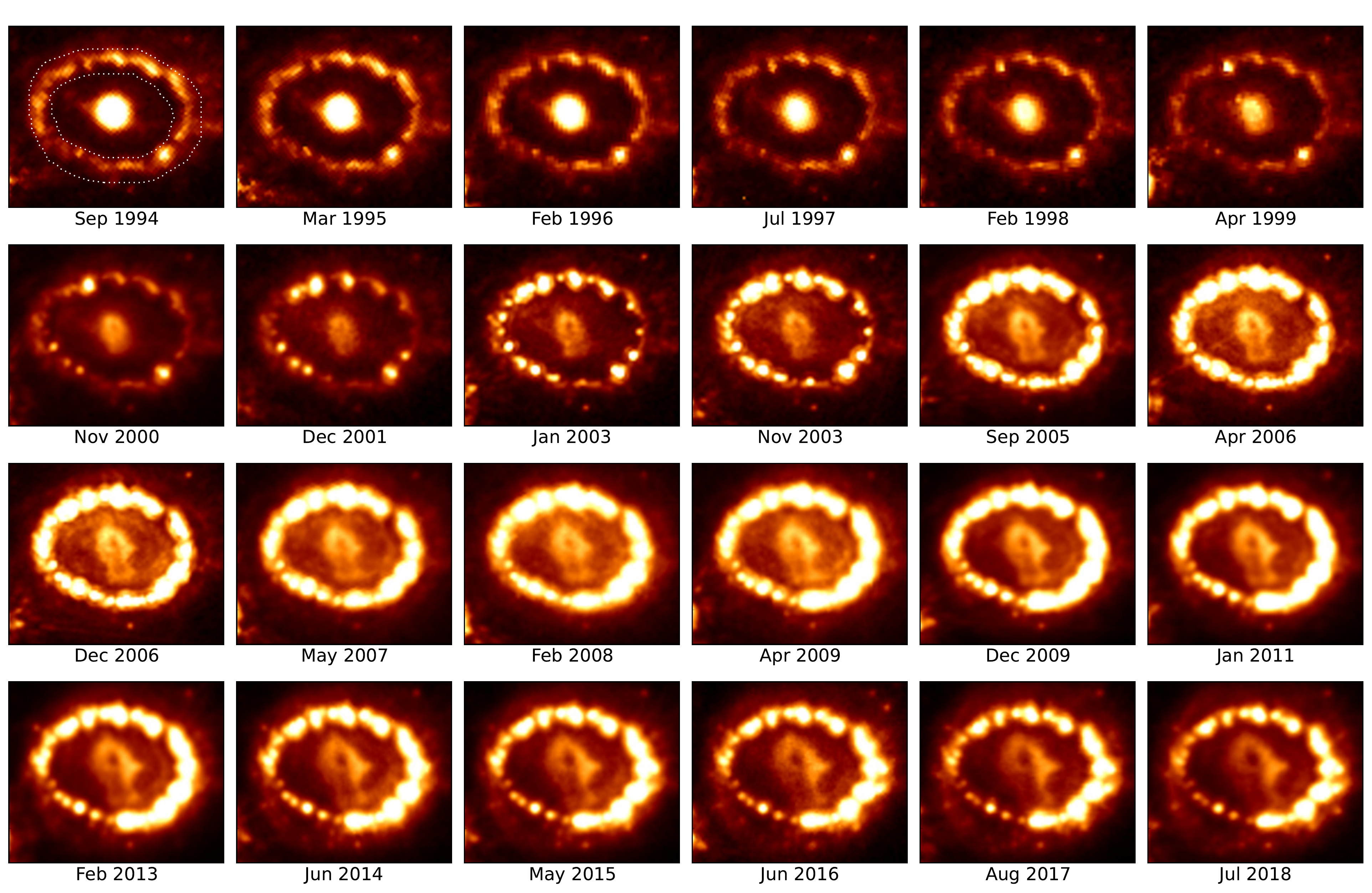}
\caption{Selection of images showing the evolution of SN~1987A in the R band.  A mask which reduces the brightness of the ER by a factor of 14 has been applied to the images (shown by the dotted lines in the top left panel). The color map scales linearly with the surface brightness both inside and outside the mask. The spot at 5 o'clock that can be seen in the ER already in the first observation is a star. The emission in the lower left corners is due to Star~3 (see Figure~\ref{fig:3ring}). North is up and east is left. The FOV is $2\farcs{1} \times 1\farcs{8}$. An animation of the evolution is available as online material.  \\\label{fig:allred}}
\end{figure*}

\begin{figure*}[t]
\plotone{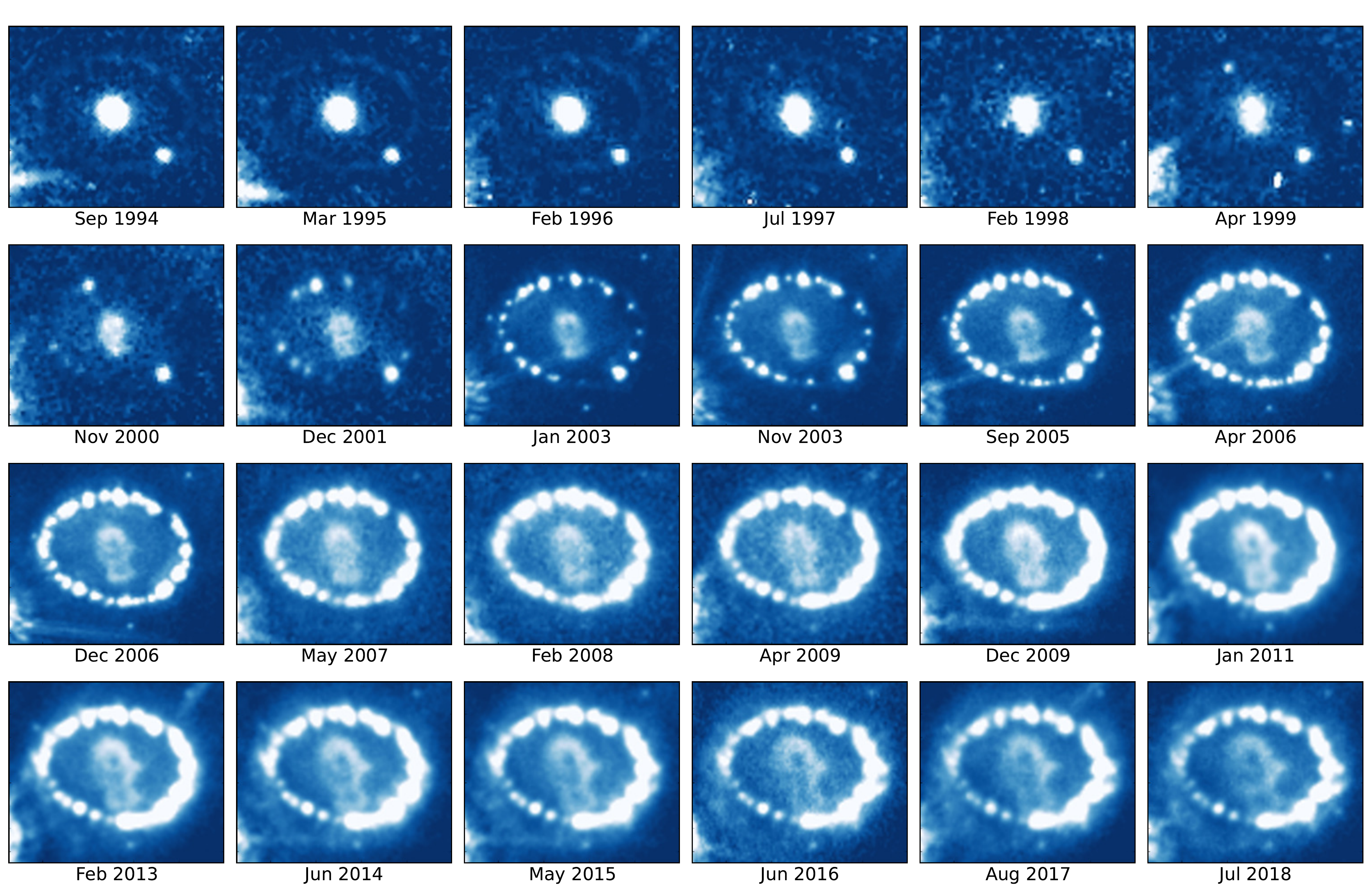}
\caption{Same as Figure~\ref{fig:allred}, but for the B band.  \\\label{fig:allblue}}
\end{figure*}

\begin{figure*}[t]
\plotone{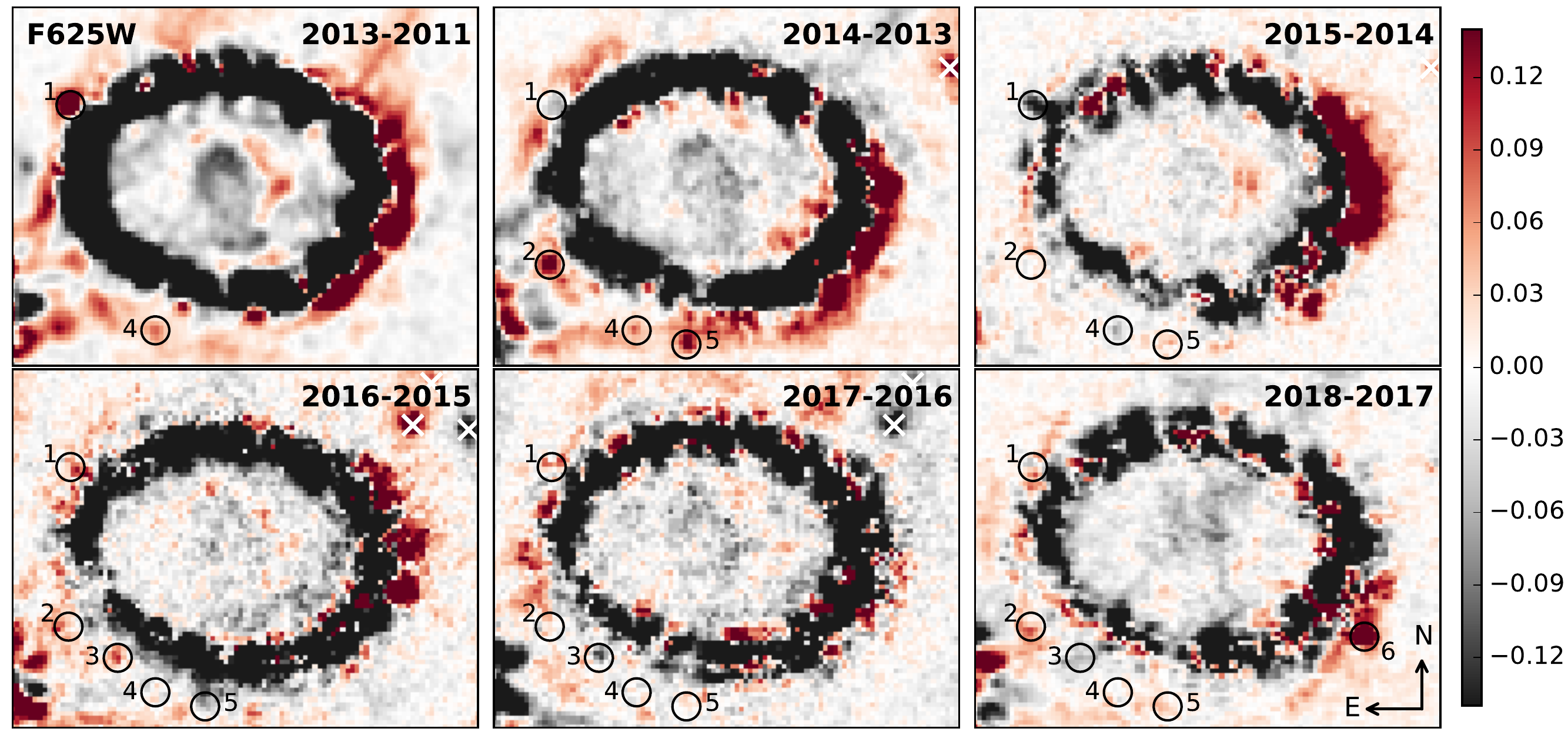}
\caption{WFC3/F625W differences images. All images except the first one are separated by one year. The spots marked by white crosses to the north-west of the ER are ghost reflections. There are also diffraction spikes and other residuals from Stars 2 and 3 in the north-west and south-east corners, respectively (see Figure~\ref{fig:3ring}). We provide flux measurements for the six labelled spots outside the ER  in Section~\ref{sec:spots}. The numbering of the spots is done anti-clockwise from north.  The unit for the color bar is counts~s$^{-1}$ per $0\farcs{025}$ pixel. The FOV is $2\farcs{5} \times 1\farcs{9}$.\\\label{fig:rdiff}}
\end{figure*}

\begin{figure*}[t]
\begin{center}
\resizebox{120mm}{!}{\includegraphics{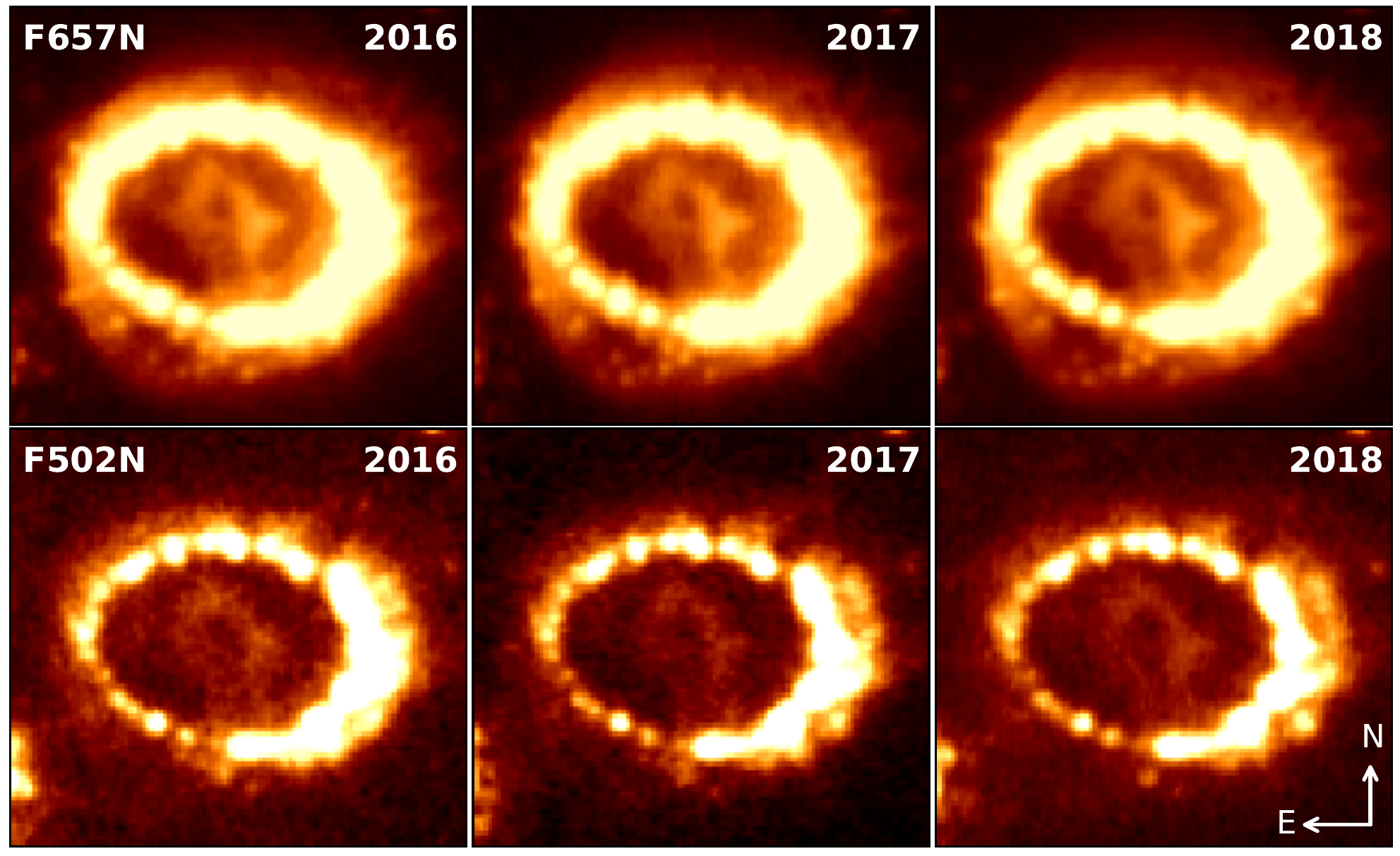}}
\caption{WFC3 narrow-band images. The top row shows F657N, for which all emission sites are dominated by H$\alpha$. The bottom row shows F502N, where the ER emission is dominated by [O~III] and the ejecta emission by Fe~I~$\lambda 5007$.  The emission in the south-east corners is due to Star~3 (see Figure~\ref{fig:3ring}). The images have been scaled by an asinh function in order to highlight the weak emission outside the ER. The FOV is $2\farcs{7} \times 2\farcs{4}$.\\\label{fig:narrow}}
\end{center}
\end{figure*}

\begin{figure*}[t]
\plotone{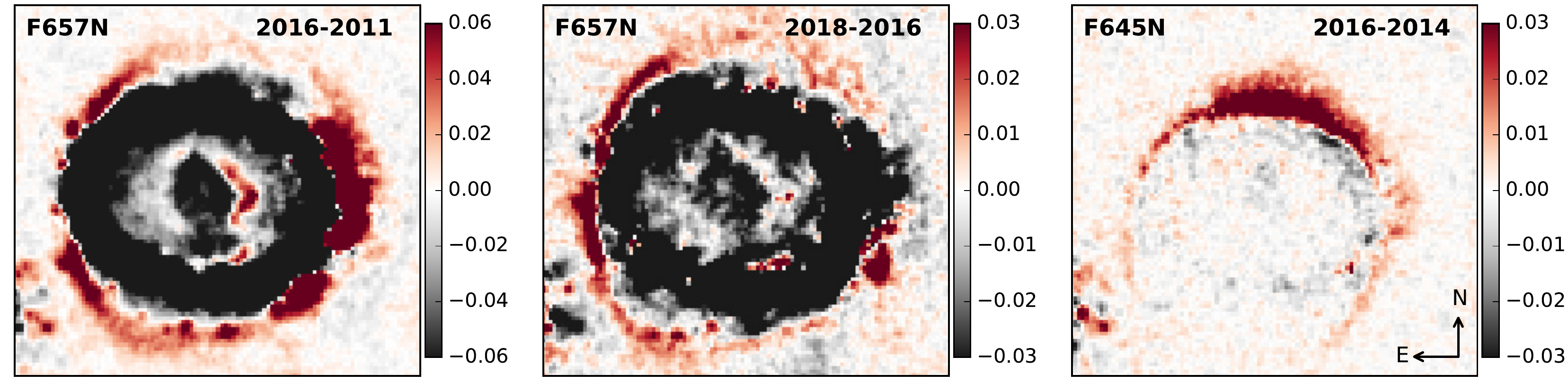}
\caption{Differences of narrow-band images in H$\alpha$. The left and middle panels show  differences in F657N between  $10,700-8700$~days ($2016 - 2011$) and $11,500-10,700$~days ($2018 - 2016$), respectively. This filter covers H$\alpha$ in the velocity range $[-3500,+3600]$~\kms. The right panel shows the difference in F645N (blueshifted H$\alpha$ in the range $[-7900,-2700]$~\kms)\  between $10,700-10,000$~days ($2016 - 2014$). The residuals in the south-east corners are due to Star~3 (see Figure~\ref{fig:3ring}). The unit for the color bar is counts~s$^{-1}$ per $0\farcs{025}$ pixel. The FOV is $2\farcs{7} \times 2\farcs{4}$.\\\label{fig:hadiffs}}
\end{figure*}
\begin{figure*}[t]
\plotone{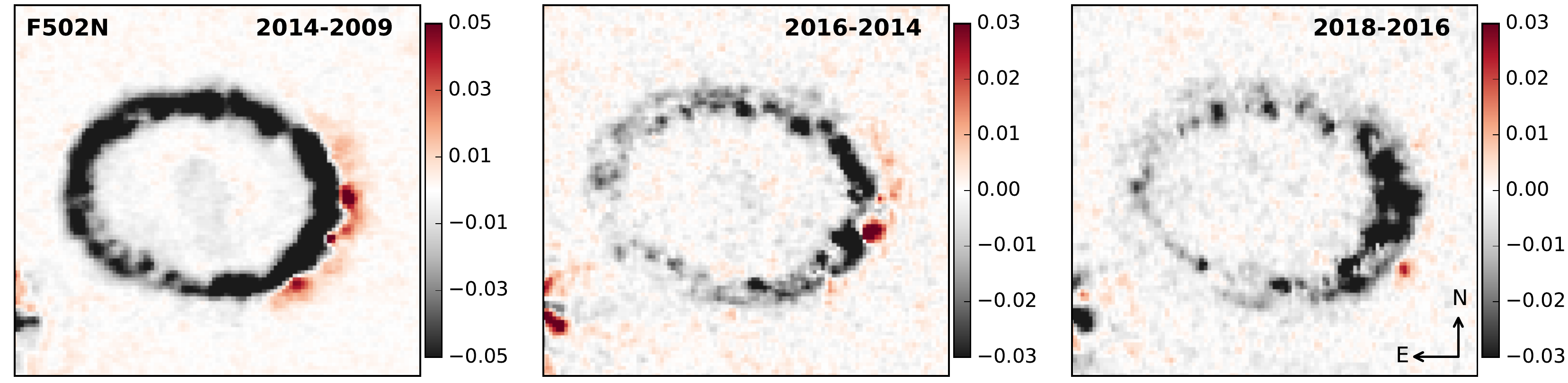}
\caption{Differences of F502N [O III] images. The epochs from left to right are $10,000-8300$~days ($2014 - 2009$), $10,700-10,000$~days ($2016 - 2014$) and $11,500 - 10,700$~days ($2018 - 2016$). The residuals in the south-east corners are due to Star~3 (see Figure~\ref{fig:3ring}). The unit for the color bar is counts~s$^{-1}$ per $0\farcs{025}$ pixel. The FOV is $2\farcs{7} \times 2\farcs{4}$.\\\label{fig:o3diffs}}
\end{figure*}

Figures  \ref{fig:allred} and \ref{fig:allblue} show a selection of images in the R and B bands, respectively, illustrating the evolution of the remnant between 2500 and 11,500 days (1994 and 2018).  A mask which reduces the brightness of the ER by a factor of 14 has been applied to the images, which makes it possible to simultaneously display the morphology of the ER and ejecta. The boundaries of the mask (plotted in the first panel of Figure~\ref{fig:allred}) were smoothed with a Gaussian. The images show a strong evolution in brightness and morphology of both the ER and the ejecta.  The most recent evolution is illustrated in Figure~\ref{fig:rdiff}, where we show difference images in the R band for all pairs of consecutive observations since 8700 days (2011). The first of these difference images (where the second observation is from 9500 days in 2013) shows the first evidence of faint spots outside the ER. This is followed by the appearance of additional spots and diffuse emission outside the ER. At the same time the main ER is fading, with the exception of the outer part of the western side. Difference images in the B band show the same trends but with lower signal-to-noise ratio (S/N).

Figure~\ref{fig:narrow}  illustrates the evolution in the  narrow F657N and F502N filters since 10,700~days in 2016 (previous narrow band images are shown in \citealt{Fransson2015} and \citealt{France2015}). These filters include all the emission from the ER, but the highest ejecta velocities and a significant part of the RS emission are excluded (see Section~\ref{sec:optem} and Table~\ref{tab:lines}). The evolution in F657N is similar to that observed in the R and B bands, with the main difference being that the diffuse emission outside the ER is even more prominent in F657N. This is seen most clearly in the  left and middle panels of  Figure~\ref{fig:hadiffs}, which show difference images in this filter between the epochs  $10,700-8700$~days ($2016 - 2011$) and $11,500-10,700$~days ($2018 - 2016$), respectively. There are no observations in F657N between 2011 and 2016. The difference images reveal a bright rim of diffuse emission outside the ER. The rim extends toward the ORs (cf.~Figure~\ref{fig:3ring}) rather than having the same shape as the ER, which suggests that the emission may originate from material connecting the rings (discussed further in Section~\ref{sec:disc-diffuse} below). 

The right panel of Figure~\ref{fig:hadiffs} shows a difference image in F645N for the time period $10,700-10,000$~days ($2016 - 2014$). This filter probes continuum emission from the ring and blueshifted  H$\alpha$ in the range $[-7900,-2700]$~\kms. The image from 2014 has been presented in \cite{Fransson2015} and \cite{France2015}, where strong diffuse emission from the northern part of the ER and streaks of emission extending along the sides of the ER were noted. This was interpreted by \cite{France2015} as emission from the RS and high-latitude material. The difference image in Figure~\ref{fig:hadiffs} shows that this emission has increased, while the continuum emission from the hotspots has decreased. The diffuse emission is most likely related to part of the diffuse emission seen in F657N, as discussed in Section~\ref{sec:disc-diffuse}.

The evolution of the [O III] emission clearly differs from that observed in H$\alpha$, as illustrated by the F502N images in Figure~\ref{fig:narrow} and the difference images in  Figure~\ref{fig:o3diffs}. The epochs for the difference images reflect the availability of observations, covering the time periods  $10,000-8300$~days ($2014 - 2009$), $10,700-10,000$~days ($2016 - 2014$) and $11,500 - 10,700$~days ($2018 - 2016$). The rim of diffuse emission outside the ER is not seen in these images. It is also noteworthy that only two new spots are seen outside the ER (one to the south and one to the south-west), compared to the dozen spots observed in F657N and in the R band. The  difference images also show a curious double structure of the ER itself, which becomes more apparent with time. This can also be seen directly in the western side of the ER in the images in Figure~\ref{fig:narrow}.

\section{Light curves}
\label{sec:lightcurves}

\begin{figure}[t]
\plotone{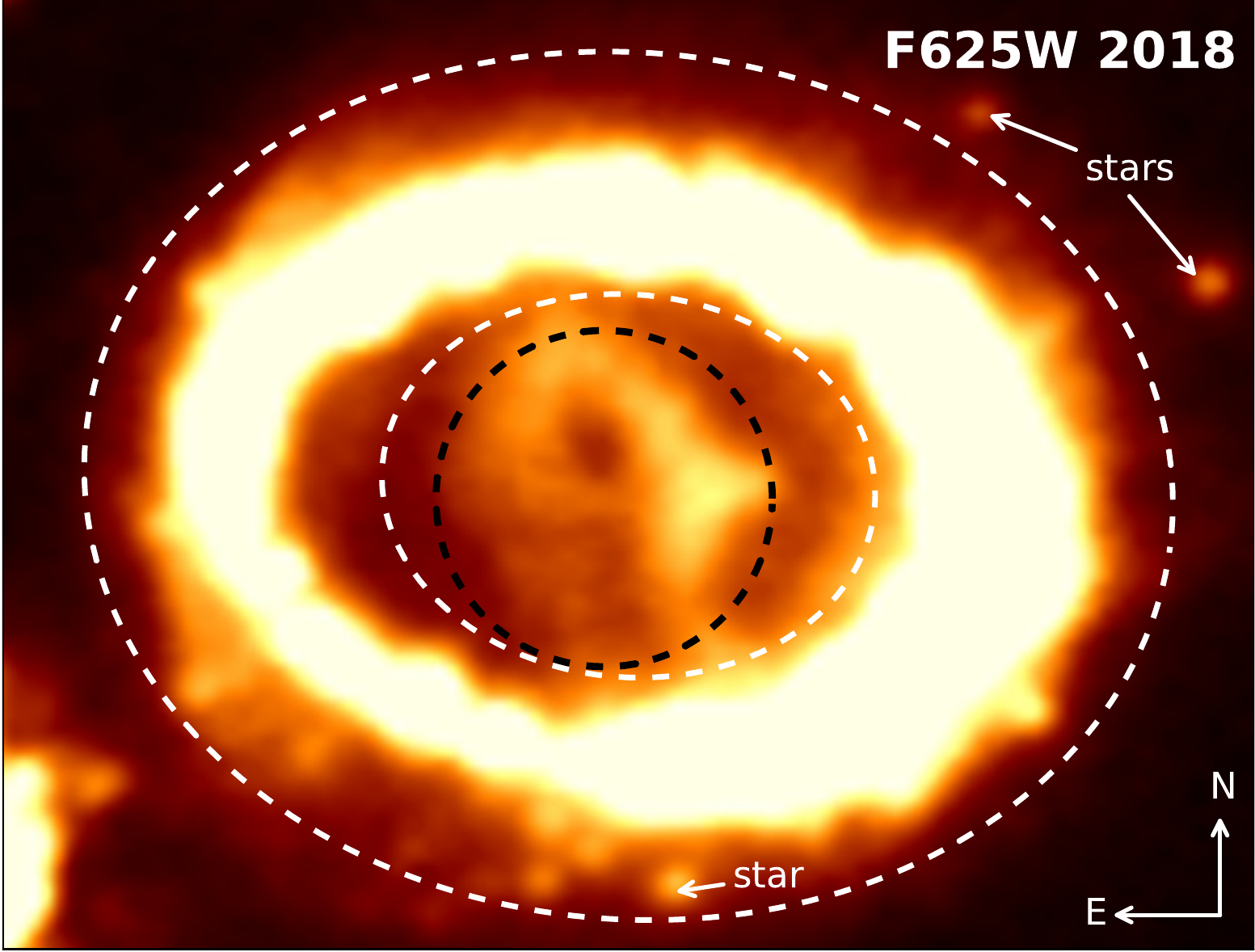} 
\caption{WFC3/F625W image from 11,500~days (2018) together with the apertures used to produce the light curves. The dashed, white and black lines show the apertures for the ER and the ejecta, respectively. The aperture for the ejecta was expanded over time in order to always cover the same part of the inner ejecta. The image has been scaled by an asinh function in order to highlight the weak emission outside the ER. The FOV is $2\farcs{7} \times 2\farcs{1}$.\\\label{fig:apertures}}
\end{figure}

\begin{figure*}[t]
\plottwo{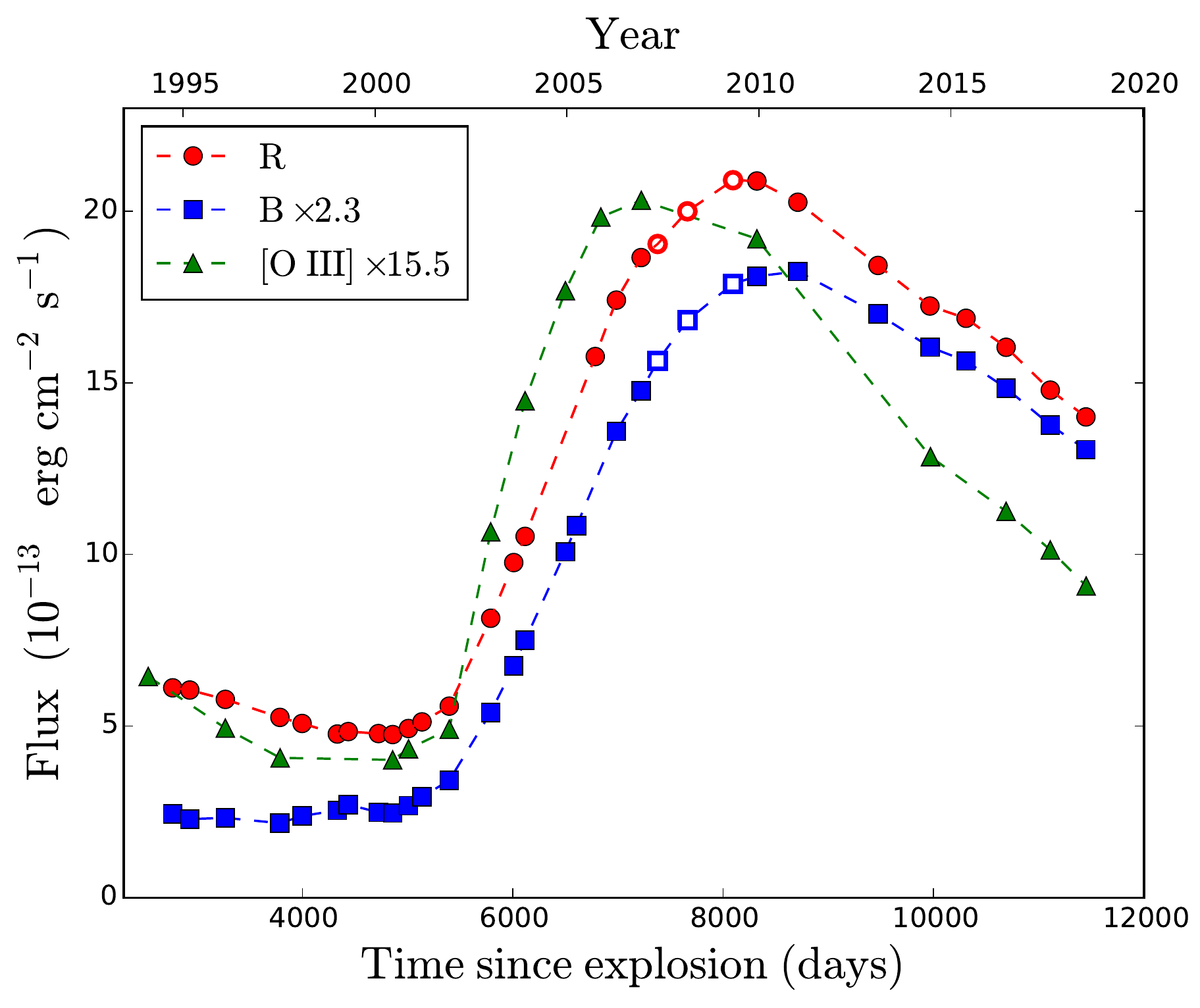} {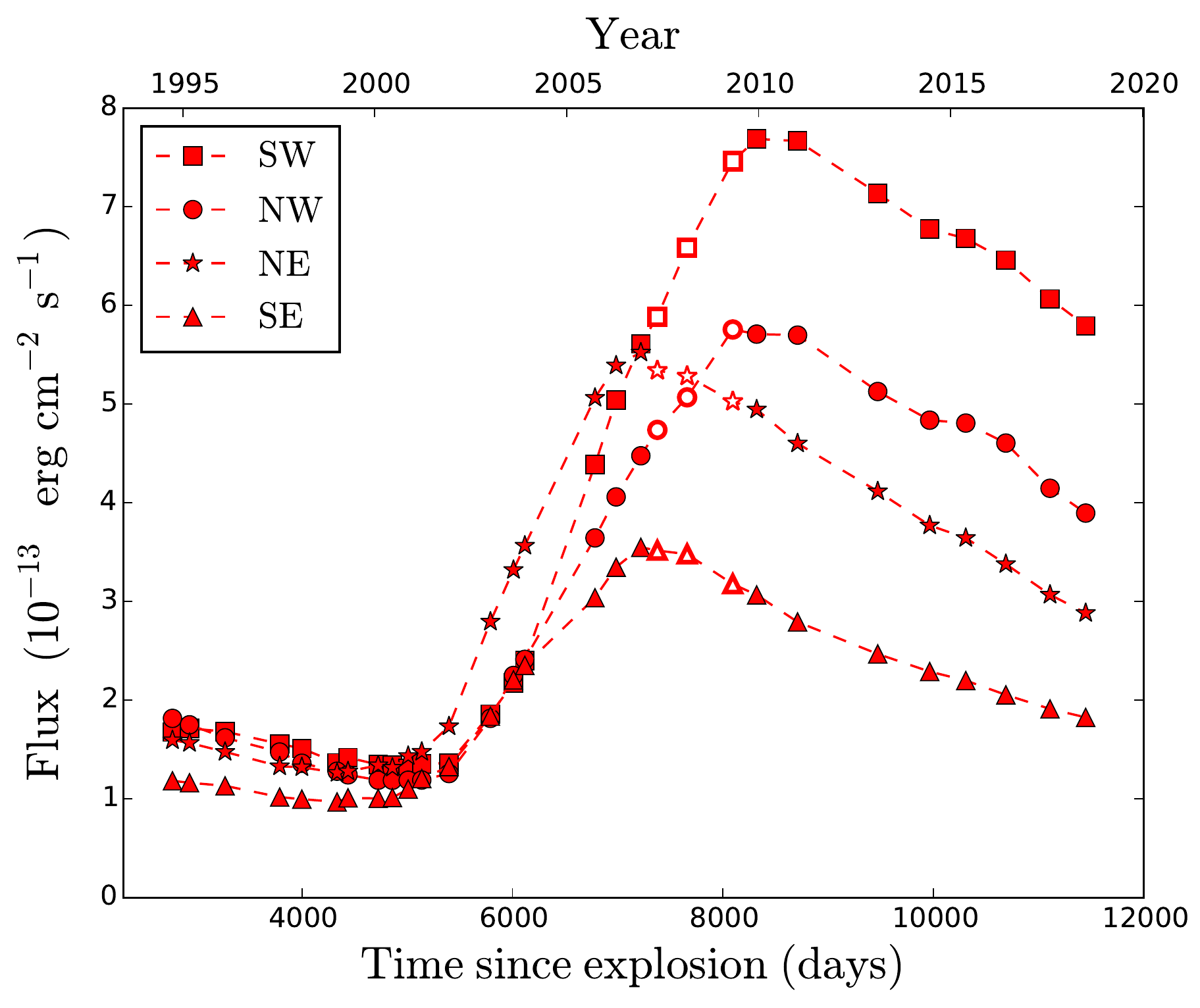}
\caption{Left: Light curve of the ER produced using the aperture in Figure~\ref{fig:apertures}. Right: Light curve in quadrants for the R band (the B band and F502N show the same trends). The statistical error bars are always similar in size or smaller than the plot symbols. The late WFPC2 fluxes (indicated by open symbols) are uncertain because the images are significantly affected by CTE losses. The fluxes in these observations have been increased by 8 and 4\% for R and B, respectively.  \\ \label{fig:ringlc}}
\end{figure*}

\begin{figure*}[t]
\plottwo{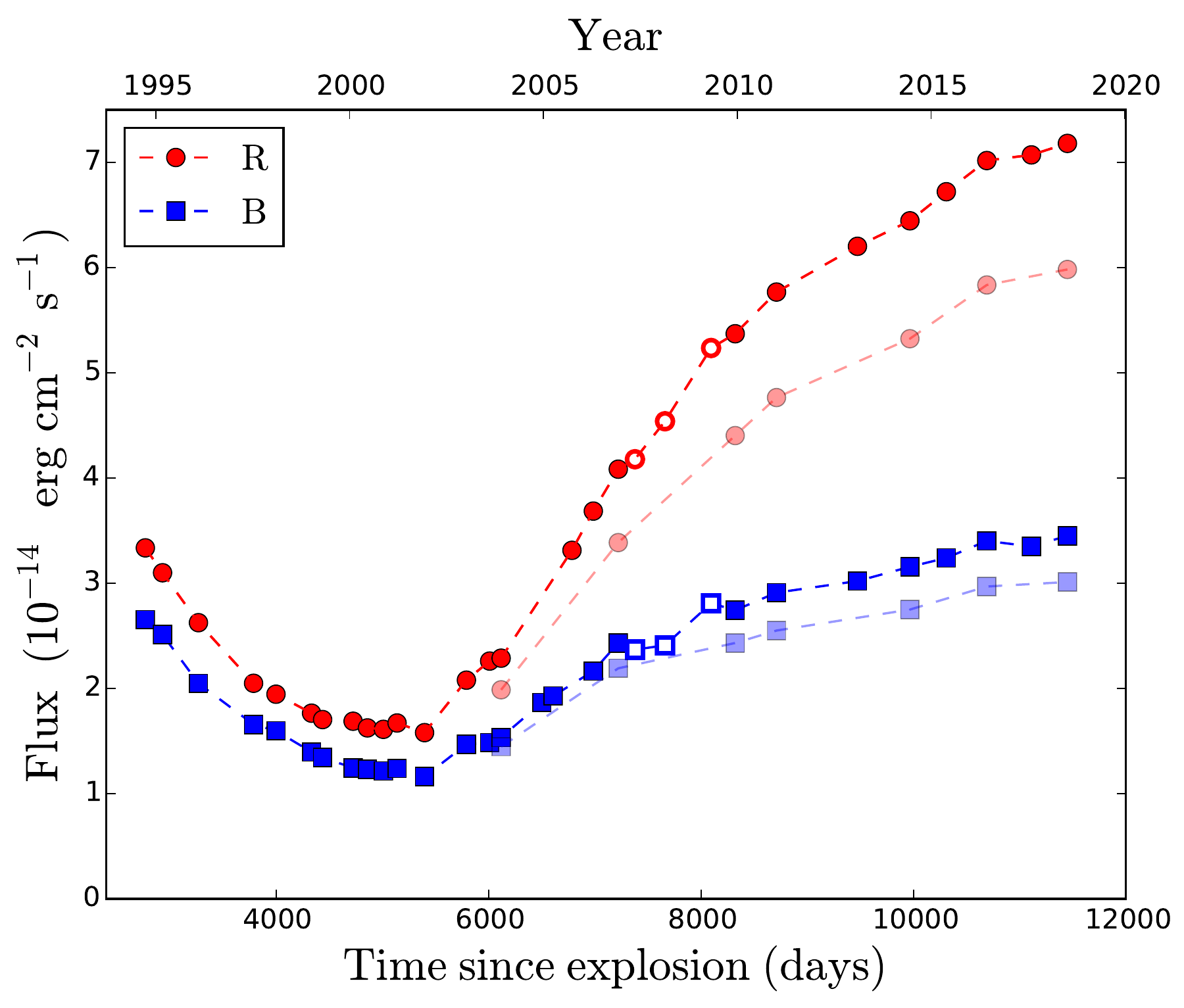} {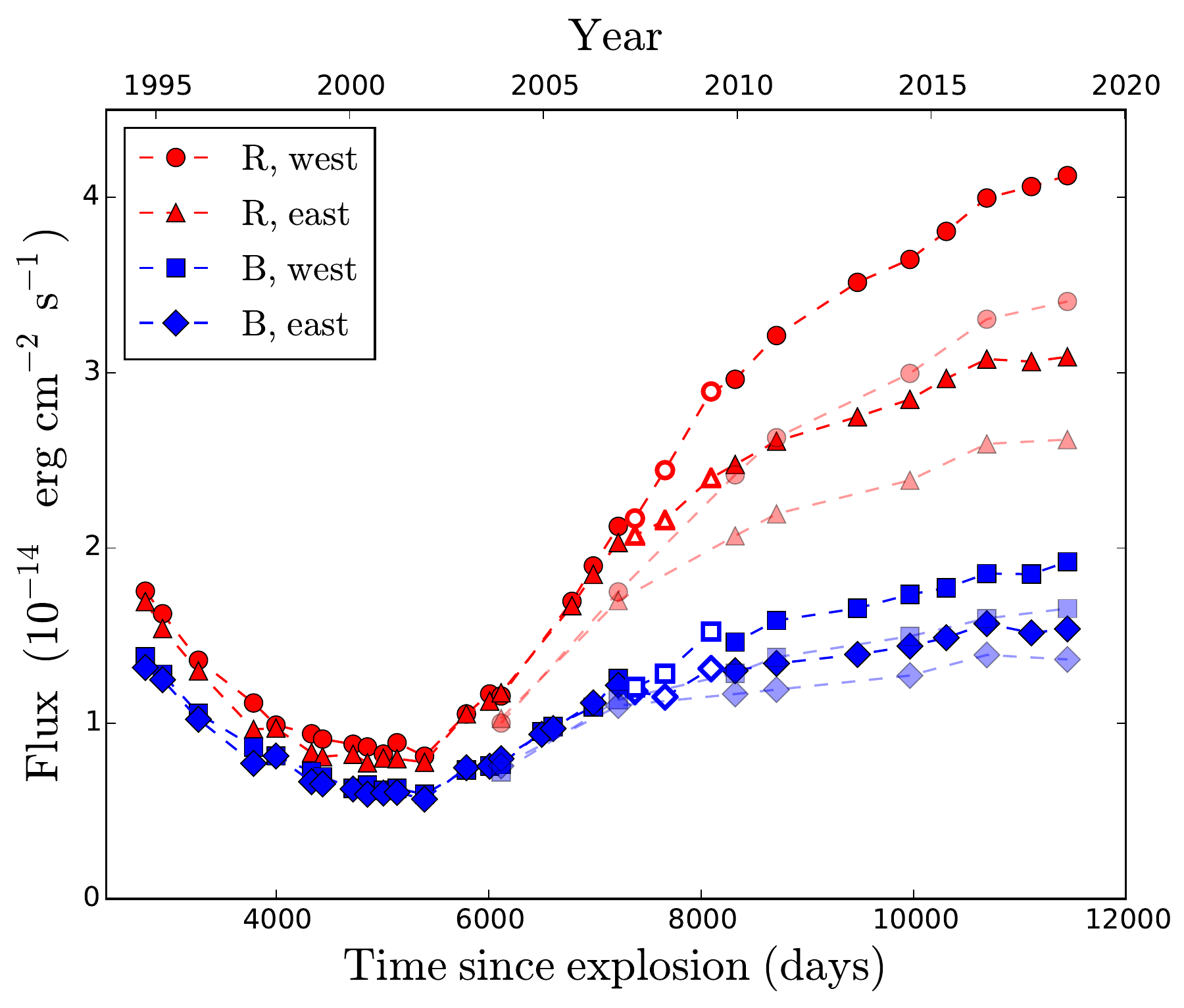}
\caption{Left: Light curve of the inner ejecta. The flux was measured in an expanding, circular aperture with radius $2800$~\kms\  in the plane of the sky (see Figure~\ref{fig:apertures}). Right: Same, but separately for the eastern and western parts of the ejecta. Note the brightening of the western ejecta at late times. The statistical error bars are always similar in size or smaller than the plot symbols. The light colored (semi-transparent) points have been corrected for scattered light from the ER using the models described in Appendix~\ref{app:tinytim}. The late WFPC2 fluxes are uncertain because the images are affected by CTE losses, but no corrections have been applied (see text for details). These points are plotted with open symbols.  \\ \label{fig:ejectalc}}
\end{figure*}

In this Section we present light curves for the ER and ejecta in the R and B bands, as well as in the F502N filter for the ER. The latter is not useful for the ejecta since it is too narrow to capture the broad lines.  The remnant has been monitored regularly in these filters since $\sim 2500$~days (1994), albeit with a more sparse sampling in F502N. The fluxes presented below  correspond to observed quantities, without corrections for extinction, but normalized to WFPC2 as described in Section~\ref{sec:obs}.  

The flux of the ER was measured in the elliptical annulus shown in Figure~\ref{fig:apertures}, chosen to include as wide an area as possible, while avoiding the ejecta and the bright stars outside the ER. However, the aperture does include the star projected on the south-west part of the ER seen in Figures~\ref{fig:allred} and \ref{fig:allblue}, as well a very faint star to the south of the ER, indicated in Figure~\ref{fig:apertures}. The resulting light curves for the three filters are shown in the left panel of Figure~\ref{fig:ringlc} and listed in Table~\ref{tab:ringlc} (Appendix~\ref{app:lc-tables}). The late WFPC2 fluxes, which are significantly affected by CTE losses (see Section~\ref{sec:obs}), were increased by 8 (4)\%  in R (B) in order to match the light curve evolution. It is notable that the light curve evolves faster in F502N, peaking at $\sim 7000$ days, compared to $\sim 8000$ days for the R and B bands.  The right panel of Figure~\ref{fig:ringlc} shows the R band light curves obtained when the elliptical aperture is split into four quadrants. This reveals large differences, as also expected from the images in Figure~\ref{fig:allred}. The south-west quadrant is more than twice as bright as the south-east one. The northern quadrants are in between these extremes, with the western part being brighter at late times. Light curves of quadrants in the B band and F502N show the same trends.

The decaying parts of all the light curves of the ER are well fitted by straight lines. Normalizing to the peak flux, $(F/F_{\rm{max}}) = c - k \times (t/10,000\ \mathrm{days})$, the slopes in the different filters are $k_{\rm{R}}=1.05\pm 0.03$, $k_{\rm{B}}=1.04\pm 0.04$ and $k_{\rm{O~III}}=1.39\pm 0.10$.  The different quadrants show that the decay is somewhat steeper on the eastern side:  $k_{\rm{R,NE}}=1.31\pm 0.03$, $k_{\rm{R,SE}}=1.25\pm 0.06$, $k_{\rm{R,NW}}=1.02\pm 0.07$, $k_{\rm{R,SW}}=0.80\pm 0.04$. The reason for this is the brightening of the outermost part of the western side of the ER around 10,000~days (2013--2015) seen in Figure~\ref{fig:rdiff}. 

The flux of the ejecta was measured in a circular aperture that was expanded with time in order to always include the same part of the freely expanding ejecta. The radius corresponds to a constant velocity of 2800~\kms\ in the plane of the sky, while the linear dimensions increase by a factor of 4.1 between the first and last observations. Figure~\ref{fig:apertures} shows this aperture in the most recent observation. The maximum size was chosen in order to avoid direct overlap with the bright emission from the ER. As a result of this limitation, part of the ejecta is outside the aperture in the south, as seen in Figure~\ref{fig:apertures}. 
   
The resulting light curves are shown in the left panel of Figure~\ref{fig:ejectalc} (and listed in Table~\ref{tab:ejectalc} in Appendix~\ref{app:lc-tables}), while the right panel shows the light curves when the expanding aperture is split into a western and eastern half. We do not apply the empirical corrections for CTE losses in WFPC2 used for the ER above, as we find that this results in discontinuities in the ejecta light curves. This difference may be explained by the fact that the background in the ejecta region is relatively high (due to the ER), which is known to reduce CTE losses. Figure~\ref{fig:ejectalc} instead indicates that smaller corrections that are different for the three epochs would be required for the ejecta. 

 As previously reported in \cite{Larsson2011}, the ejecta light curve decays until $\sim 5000$ days and then increases. With the new observations added here we note that the flux is continuing to increase, but at a gradually slower rate at later times. The eastern and western halves evolve very similarly until 7000~days, after which the western part brightens faster and remains brighter than the eastern part. The flux in the eastern half has remained approximately constant since 10,700 days.

The main uncertainty affecting the ejecta light curves is the background due to scattered light from the ER. The ER is more than an order of magnitude brighter than the ejecta and the tails of the Point-Spread Function (PSF) of the hotspots will contribute a significant flux in the ejecta region. This effect is negligible before 6000 days, but  becomes more important as the ER brightens and the aperture for the ejecta expands. To obtain a first-order estimate of this effect we construct synthetic models for the ER at seven epochs between $\sim 6100 - 11,500$~days. The details of this are described in  Appendix~\ref{app:tinytim}. From the models we can compute the predicted flux due to scattered light in the aperture of the ejecta. This amounts to $\sim 17\ (12)\%$ of the total measured ejecta flux in the R (B)-band after 8700~days, with the fading ER and expanding aperture combining to produce a roughly constant contribution. The effect of this on the light curve is shown in Figure~\ref{fig:ejectalc}. We note that the corrections do not change the conclusion that the flux in the western part of the ejecta is still increasing. 

Finally, a smaller source of uncertainty is Star~3, which is located to the south-east of the ER. Depending on the roll angle of the telescope, diffraction spikes from this star may overlap with the remnant. The only instance where this has a significant impact is at 7000~days (2006 April,  see Figure~\ref{fig:allblue}), where such a spike affects the ejecta light curve in the B band. We estimated the contribution from this as the average flux of the corresponding parts of the other diffraction spikes of Star 3. Correcting the light curve for this resulted in a $\sim 10\%$ reduction of the measured flux. The effect of the diffraction spike is negligible for the ER and for the ejecta in the R band due to the much higher fluxes relative to the spike.

\section{Expansion of the ER}
\label{sec:expansion}

\begin{figure}[t]
\plotone{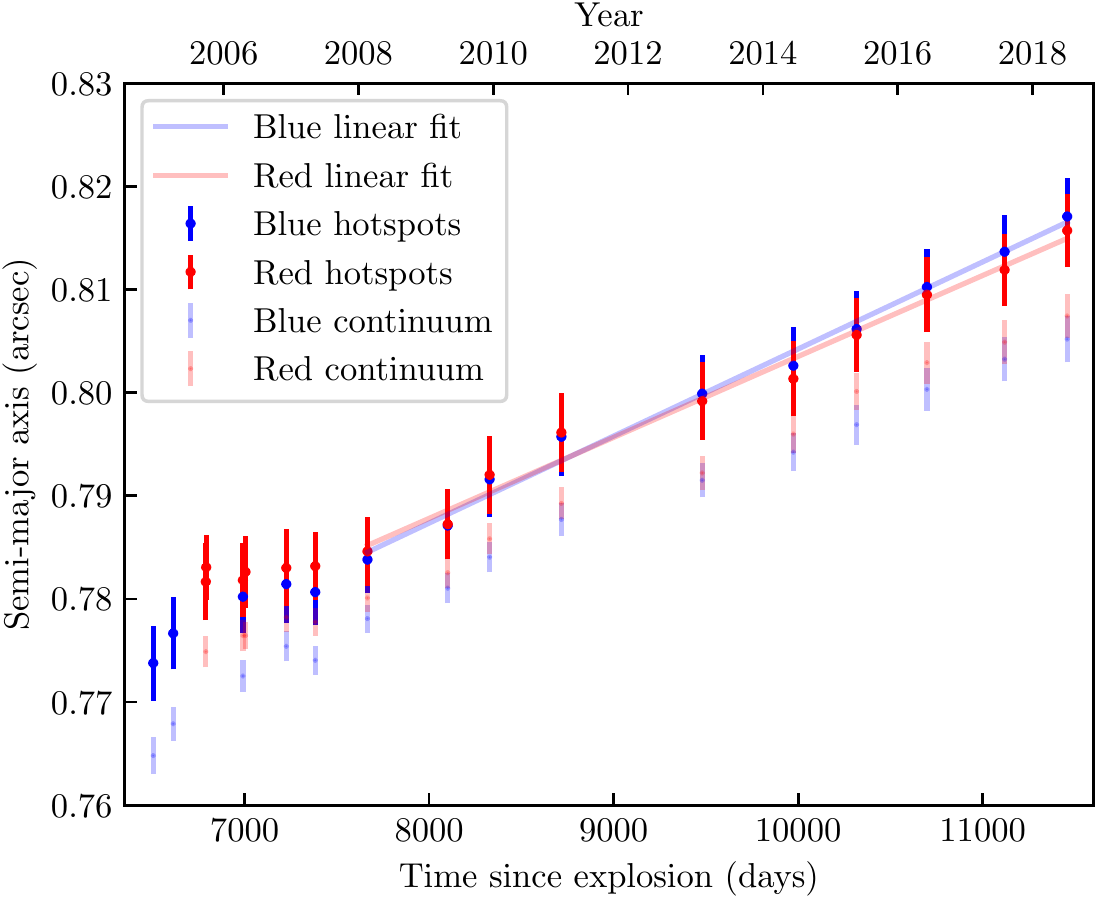}
\caption{Evolution of the semi-major axis of the ER. The lines are fitted to the hotspot measurements after 7700~days, whereas the lines fitted to the continuous elliptic annuli are omitted for visual clarity.  \\\label{fig:ringexp}}
\end{figure}

We analyzed the expansion of the ER by fitting ellipses
to the positions of the hotspots (similarly to \citealt{Fransson2015} and \citealt{Alp2018}).
We used the 26 hotspots identified in the R-band image from 7200~days (2006 December, see Figure~1
of \citealt{Alp2018}). The positions of the hotspots were determined by
fitting 2D Gaussians to this image. The hotspots in the other observations were then located by
fitting 1D Gaussians to the radial profiles defined by the position
angles for the hotspots identified in the 2006 image. This ensures that the same hotspots are identified in all
observations.

We performed all fits using the Levenberg-Marquardt algorithm and took the square root of the diagonal elements of
the covariance matrix as the standard deviations of the fitted
parameters. The assumption that the hotspots are described by a perfect ellipse
introduces a systematic uncertainty. To explore the effects of this,
we also fitted an elliptic annulus with sinusoidal intensity along the
azimuthal direction and a Gaussian radial profile. The aim of this
 is to capture all the emission from the ER rather than
treating it as a collection of point sources. The results are shown in Figure~\ref{fig:ringexp}. We find that the
two methods give similar results, but that the systematic uncertainties dominate the
error budget.

We use the temporal evolution of the semi-major axis from the fits 
to measure the expansion velocity, which should correspond to the typical velocity of the transmitted shocks in the hotspots. We
focus on observations later than 7700~days because the
evolution is less systematic at earlier times. From the fits of
ellipses to the hotspots, we find an expansion velocity of
$696\pm36$~\kms\  in the R band and $748\pm30$~\kms\  in the B band.
This can be compared to $639\pm25$~\kms\  in the R band and
$626\pm22$~\kms\  in the B band from the fits of the extended,
continuous elliptic annuli. The errors are only statistical, which
shows that the systematic uncertainties are dominating. For this reason, we take $680\pm50$~km~s$^{-1}$ as our estimate of the expansion velocity, which corresponds to the average and the sample standard deviation of the four individual values. For comparison, 
\citet{Fransson2015}  found velocities for individual hotspots in the
range 180--950~\kms\  with a mean of 540~\kms. We note that those velocities were not corrected for the inclination of the ER, whereas our measurement based on the semi-major axis represents the true velocity.

\section{Properties of the new spots}
\label{sec:spots}

Since  $\sim 9500$~days a number of new, faint spots have appeared outside the ER (Figures \ref{fig:rdiff} -- \ref{fig:o3diffs}). We note that the spots that can be seen appearing and disappearing to the north-west of the ER in the R band in Figure \ref{fig:rdiff} are due to ghost reflections of Star~2 (labeled in Figure~\ref{fig:3ring}). The reflections can be identified because similar features appear around all the bright stars in a given image and because the spots move between consecutive observations as the roll angle changes. Bright reflections in the vicinity of SN~1987A are only observed in the R band. There are also three faint background stars located near the ER (labelled in  Figure~\ref{fig:apertures}), which can be seen in all previous broad-band observations. Excluding the reflections and stars, we identify a dozen new spots associated with SN~1987A, most of which are located in the south-east.  These all fulfill the criteria that they (i) are not detected before 9500~days, (ii) are detected in at least two different filters, and (iii)  cannot be identified as reflections based on a comparison of features around bright stars in the image.    

We performed flux measurements for the six new spots labelled in Figure~\ref{fig:rdiff}. The remaining spots were either too faint and/or located too close to the ER to enable reliable measurements. Aperture photometry was performed with the IRAF/DAOPHOT package \citep{Stetson1987}, using an aperture radius of $0\farcs{075}$. The centers of the apertures were kept fixed at the positions of the spots determined by fitting 2D Gaussians to the difference images where each spot first appears. We verified that fits at other epochs did not show any evidence for significant movement of the spots with time. Aperture corrections were determined from measurements of bright isolated stars in the images. 
 
 The background was taken as the median in an annulus around each spot with inner and outer radii of $0\farcs{100}$ and $0\farcs{230}$, respectively. This method was chosen based on extensive testing, which showed that this gave the best agreement between the R band and F657N, as well as a flux evolution in agreement with that seen directly in the difference images. We stress that there is a large systematic uncertainty due to the complex background, which varies significantly over small spatial scales. These measurements therefore only provide a rough estimate of the flux levels and time evolution. We also performed PSF photometry on the six spots, which gave consistent results for the time evolution, but failed in a few cases for the faintest spots. 

\begin{figure}[t]
\plotone{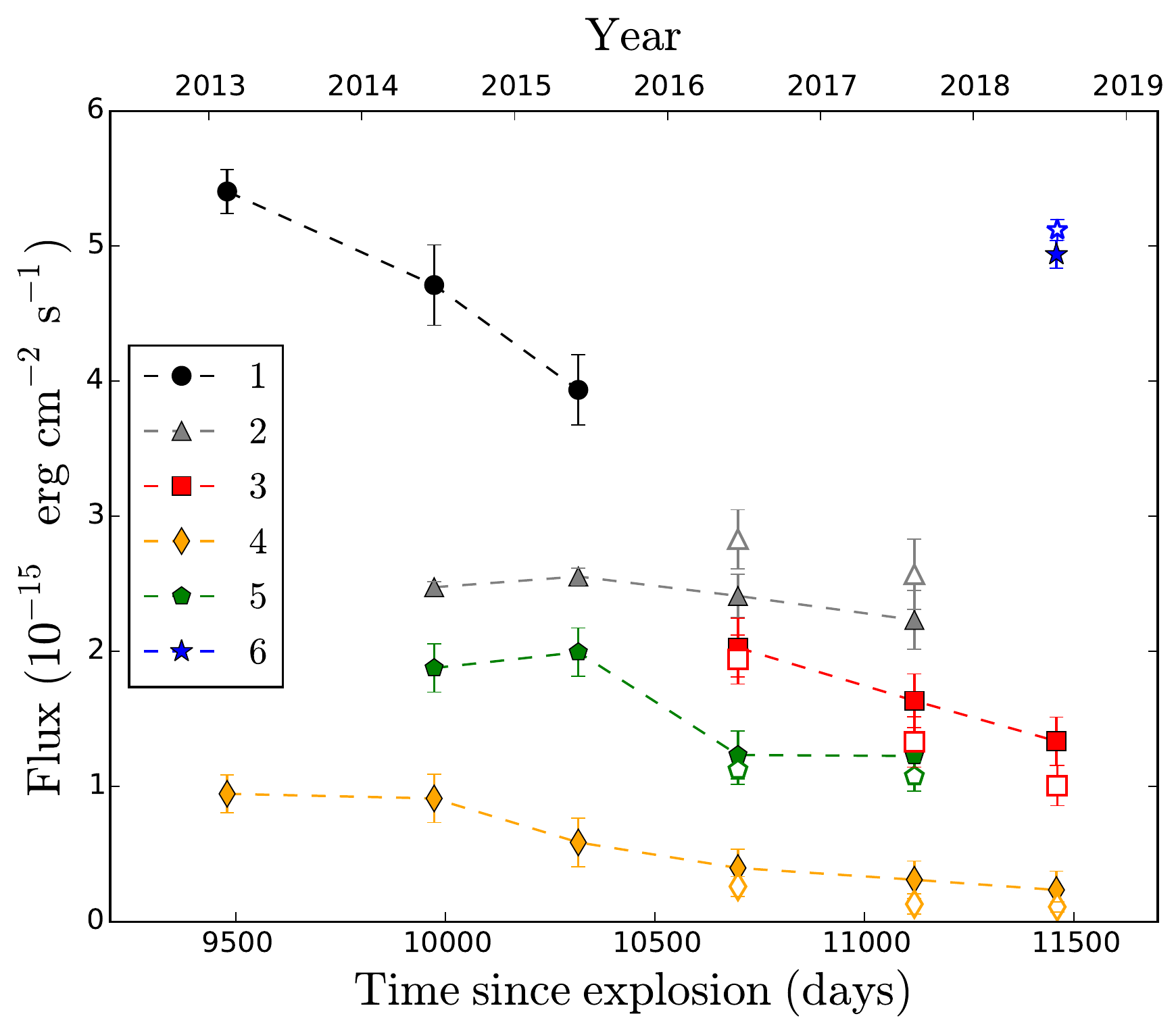} 
\caption{Light curves for six of the new spots that have appeared outside the ER. The numbers are the same as in Figure~\ref{fig:rdiff}. Filled and open symbols show fluxes in the R band and in F657N, respectively. All the observations were made with WFC3, but the R-band fluxes have been normalized to WFPC2 using the correction factor in Table~\ref{tab:corrfactors} in order to allow for comparison with the light curves of the ER. The F657N fluxes have been multiplied by a factor of 1.6.  \\\label{fig:spotlc}}
\end{figure}

The  flux measurements in the R band and F657N are shown in Figure \ref{fig:spotlc}.  The light curves of spots 1, 2 and 5 have been truncated before the last observation because the rim of diffuse emission brightens inside the apertures of the fading spots, resulting in large uncertainties.  Spots 1 and 6, which are closest to the ER, are clearly the brightest. The other spots are several times fainter and located further away. The fluxes in F657N are only $\sim$ 1.5--2 times lower than in the R band, despite the filter widths differing by a factor of 12. This is a very strong indication that the emission is due to lines (predominantly H$\alpha$ and [N II] if the spectrum is similar to that of the ER) rather than continuum. Normalizing the light curves by the peak flux and fitting  straight lines as in Section~\ref{sec:lightcurves} gives decay slopes of $k_1=3.20\pm 0.44$, $k_2=1.57\pm 0.05$ and $k_3=4.50\pm 0.07$ for the first three spots. The quoted error bars are statistical only and do not account for the systematic uncertainty in the background measurements. Spot 4 is not well described by a linear model, spot 5 only has two data points in the decaying part and spot 6 only has one flux measurement.   

\section{Discussion}
\label{sec:disc}

Below we discuss the properties of the different emission components in SN~1987A. We discuss the ER in Section~\ref{sec:disc-er}, the emission outside the ER in Section~\ref{sec:disc-outside}, and the ejecta in Section~\ref{sec:disc-ejecta}. The future evolution of all these components is discussed in Section~\ref{sec:disc-future}. \\

\subsection{Properties and evolution of the ER}
\label{sec:disc-er}

Almost all the optical emission from the ER  comes from shock-heated gas in clumps with high densities (up to $\sim 6 \times 10^4\ \rm{cm}^{-3}$ before the shock interaction, \citealt{Fransson2015}). The clumps, which are observed as hotspots in the images, are embedded in gas with lower density ($\sim 10^3\ \rm{cm}^{-3}$, e.g.~\citealt{Mattila2010}). The first appearance of  a hotspot can be traced back to $\sim 3000$ days \citep{Sonneborn1998,Lawrence2000}. This hotspot was initially very faint, but by  $\sim 4000$~days (1998)  it is clearly seen in the north-east part of the ER in Figure~\ref{fig:allred}. By  $\sim 6000$ days (2003), hotspots had appeared all around the ER. The ER continued to brighten in the R and B bands until 8000~days and then started to fade, as  reported in \cite{Fransson2015}. Here we have extended the broad-band light curves by four years and also presented the light curve for the F502N [O~III] filter. This  shows that the ER continues to fade linearly with time and that the [O~III] emission evolves faster than the R and B bands, peaking $\sim 1000$~days earlier and also taking  $\sim 1000$~days less to reach half the peak flux (Figure~\ref{fig:ringlc}).

To understand the more rapid evolution of the [O~III] emission, we first consider the recombination time scale, $t_{\rm rec}=1/(\alpha_{\rm R}(T_{\rm e}) n_{\rm e} )$, where $\alpha_{\rm R}(T_{\rm e})$ is the recombination rate as a function of the electron temperature, $T_{\rm e}$, and $n_{\rm e}$ is the electron number density. This time scale is short, although it generally increases for lower ionization stages. The temperature of the gas from which the [O III]  emission originates is $\sim 10^5$~K (if in coronal equilibrium,  \citealt{Nahar1999}). The recombination rate for O$^{2+}$ increases from $2.5\times 10^{-12} $~cm$^{3}$~s$^{-1}$ at $10^4$~K  to  $1.33\times 10^{-11}$~cm$^{3}$~s$^{-1}$ at $\sim 2\times 10^5$~K. The density in the compressed shocked gas is $10^6 - 10^7~\ccm$ \citep{Groningsson2008a}, giving a recombination time scale for  O$^{2+}$  of $t_{\rm rec}=1.2 (\alpha_{\rm R} /10^{-11}~{\rm cm}^{3}~{\rm s}^{-1})^{-1} (n_{\rm e}/10^6 {\rm ~ cm^{-3}})^{-1}$~days.  For comparison, the observed time scale for the [O III] emission to decay to 50 \% of the peak flux is $\sim 3600$~days (Figure~\ref{fig:ringlc}). It is therefore highly likely that the observed decay represents a real decrease in the shocked mass per unit time, rather than slow recombination. Although the recombination rate of H$^{+}$ is slower by more than one order of magnitude, the recombination is still fast and the difference in decay time for [O~III] and the Balmer lines (which make major contributions to the R and B bands) is unlikely to be caused by different recombination rates. 

A more likely explanation is that the [O III] emission reflects the instantaneous mass flux of shocked, cooling gas, while the emission from H I and other lower ionization stages reflects the accumulated mass of singly ionized gas behind the more rapidly cooling gas. The amount of ionized gas behind the cooling region is determined  by a balance between recombination and the ionizing extreme UV and soft X-ray emission from the shock \cite[e.g.,][]{Allen2008}.  The  slow evolution of the X-ray flux \citep{Frank2016} compared to the much faster decrease of the narrow lines from the unshocked gas in the ER (K. Migotto et al., in preparation) demonstrates that the ionizing flux is decreasing more slowly than the amount of unshocked gas

The [O III] emission from the ER also exhibits a kind of double structure, which is seen in the images after $\sim 10,000$~days in 2014 (Figures~\ref{fig:narrow} and \ref{fig:o3diffs}). The inner part  matches the hotspots of the ER observed in other filters, while the outer part can be described by an ellipse with the same eccentricity and orientation, but a semi-major axis that is $\sim 0\farcs{16}$ larger at 11,500~days. The outer part is fainter and mainly composed of diffuse emission. Inspection of the H$\alpha$ images reveal a corresponding, but very faint, structure  south-west of the ER only (hidden by the bright emission from the hotspots in Figure~\ref{fig:narrow}). This difference in morphology between the filters may be  related to the fact that the [O~III] emission reflects the immediate energy input from the shocks. If the emission from the outermost part of the ER is due to newly shocked gas, the H$\alpha$ region has not yet had time to build up. 

A plausible explanation for  the outer structure is that it is due to emission from material swept up by the blast wave moving through gas with lower density between the clumps. The velocity of the blast wave is $1850 \pm 100$~\kms\ (measured from X-rays, \citealt{Frank2016}). Assuming this velocity, the time scale to produce the observed separation between the two structures is 7400~days, which is consistent with  the start of the interaction as signaled by the appearance of the first hotspots around 4000~days. Another possible explanation for the double structure is that the ER was formed by two different ejection events. Assuming that the outer part is in the same plane and has the same expansion velocity as the inner one (10~\kms,~\citealt{Crotts2000}), the inferred time difference between the ejections is $\sim 4000$~years. We find this scenario less likely given that the ER and ORs have velocities that are consistent with a common ejection event.  A third option is that both parts of the ER were ejected around the same time, but with different velocities. There is no evidence for multiple velocity components in the spectra of the ER, but we note that a faster and fainter component may be hard to detect. 

The fact that the emission from the hotspots has stopped increasing indicates that the blast wave has left the ER and is no longer sweeping up new high-density material.  At the same time, the dense clumps are  most likely being destroyed by the shocks, as discussed in \cite{Fransson2015}. A change in the evolution of the ER has also been observed at other wavelengths. The mid-IR emission, which is due to shock-heated dust, peaked around 8500~days and shows a similar evolution as the optical light curves \citep{Arendt2016}. The shocks in the hotspots also produce soft X-ray emission, which can be described by a $\sim 0.3$~keV collisional ionization equilibrium plasma \citep{Frank2016}. The $0.5-2$~keV light curve has leveled off at a constant value at 9500~days and the images of the softest X-ray emission show a similar morphology as the optical, supporting an association with the optical hotspots \citep{Frank2016}. By contrast, both the hard X-ray and radio emission continue to increase and are brighter in the east \citep{Frank2016,Cendes2018}. This emission is thought to originate from the lower-density material ($\sim 10^3\ \rm{cm}^{-3}$) located between the clumps  and at higher latitudes. Such a low-density component is also required in models for the optical emission from the unshocked gas in the ER (e.g., \citealt{Groningsson2008a,Mattila2010}). Radio observations show that the shock velocity started to increase around 9300 days \citep{Cendes2018}, supporting the scenario that the blast wave has left the ER. Interestingly, the recent evolution of the ER may also be associated with the emergence of gamma-ray emission \citep{Malyshev2019}.

By fitting ellipses to the ER in the time series of {\it HST} R- and B-band images we find that it is expanding at $680\pm50$~\kms. This number is not affected by the inclination of the ER toward the observer, and should thus correspond to the typical velocity of the transmitted shocks in the clumps. This is compatible with the range of shock velocities observed in high-resolution spectra, although a direct comparison is complicated by the inclination and the fact that the emission is a superposition of oblique shocks of different velocities (K. Migotto et al., in preparation). The spectra also show that the highest shock velocities increase with time as faster shocks have time to cool (\citealt{Groningsson2008b}, K. Migotto et al., in preparation). 

The fits to the ER give a semi-major axis of $\sim 0\farcs{82}$ in the last epoch. This is smaller than the $1\farcs0$ radius estimated in radio \citep{Cendes2018}, as expected if the latter emission originates from material interacting with faster shocks. Interestingly, the semi-major axis of the outer part of the ER in [O~III]  is similar to the size in radio, supporting the scenario that the [O~III] emission originates from the blast wave propagating between the clumps. However, the  [O~III]  likely originates close to the plane of the ER, while the radio is well described by a torus model with a current opening angle of $\sim 30^{\circ}$ \citep{Cendes2018}.  For X-rays,  \cite{Frank2016} report a semi-major axis of $\sim 0\farcs{83}$ in the full band and  $\sim 0\farcs{75}$ in the $0.3-0.8$~keV band. The latter is smaller than the size of the optical ER, which is not expected if the soft X-ray emission arises from the hotspots. This may be due to systematic uncertainties associated with the significantly lower resolution in X-rays.

\subsection{Emission outside the ER}
\label{sec:disc-outside}

In this section we discuss the properties and origin of the emission outside the ER. We propose that this emission is due to fast ejecta interacting with high-latitude material that extends from the ER toward the ORs. In this scenario the emission from the new spots is due to slow shocks propagating in dense clumps embedded in the high-latitude material. The shocked lower-density gas between the clumps is not expected to produce optical emission. However, the interaction will also give rise to a RS, which can explain the diffuse emission outside the ER. Specifically, the H~I in the ejecta will be excited when reaching the RS and subsequently produce H$\alpha$ emission at the velocity of the freely expanding ejecta. In Section~\ref{sec:disc-rs} below, we start by briefly discussing the spectra of the RS in SN 1987A. The RS extends well outside the ER and provides information about the highest velocities of interacting ejecta. This is followed by a discussion of  the properties of the new spots and diffuse emission in Sections~\ref{sec:disc-spots} and \ref{sec:disc-diffuse}, respectively.

\subsubsection{The reverse shock}
\label{sec:disc-rs}

\begin{figure}[t]
\plotone{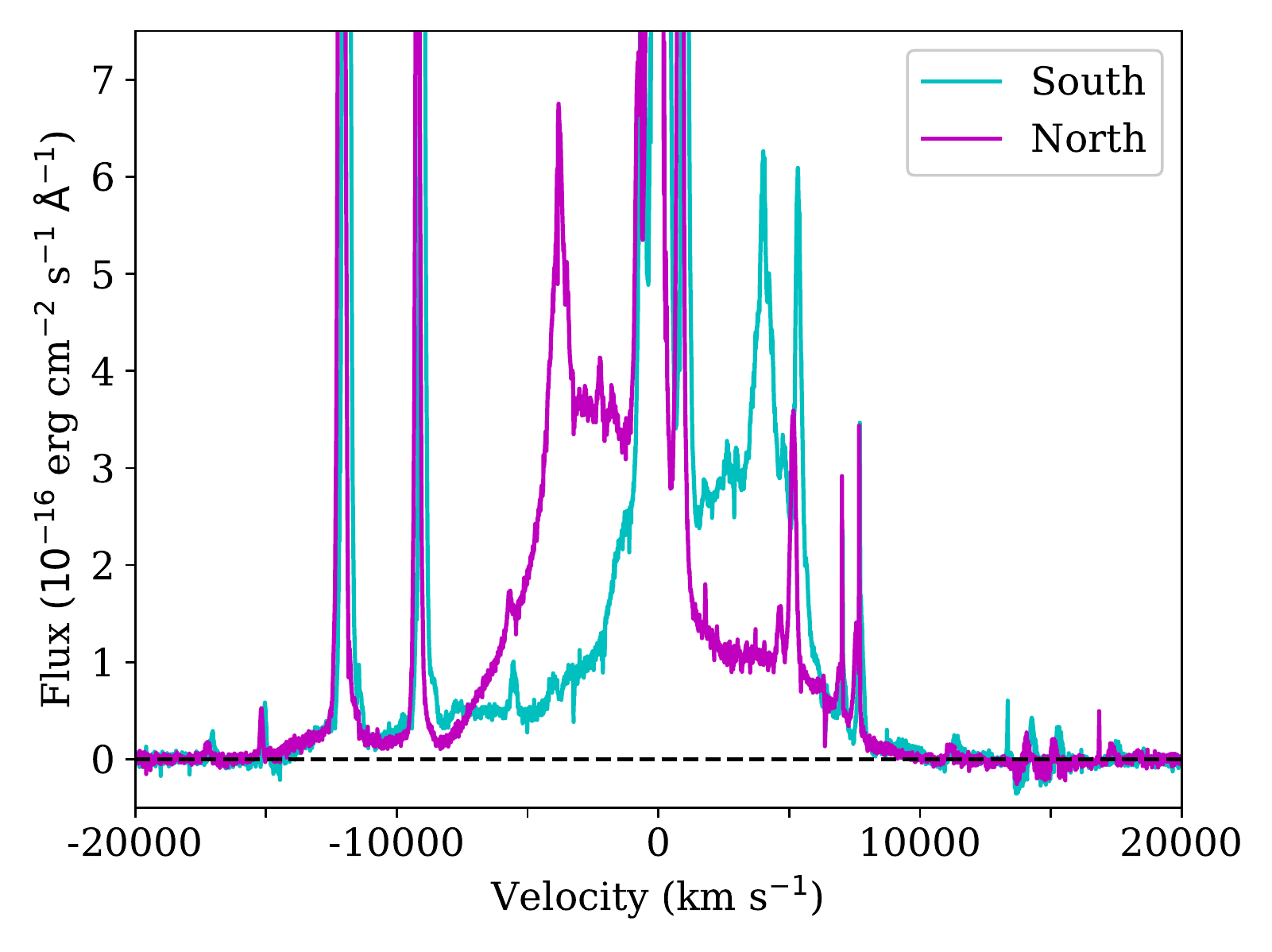}
\caption{VLT/UVES H$\alpha$ line profiles  from 11,200~days (2017). Spectra from the northern and southern halves of the slit are shown in magenta and cyan, respectively.  The narrow lines  are from slow shocks into the ER, while the extremely broad profiles extending to $\sim 9000$~\kms\ are from the RS.  The features at $\sim \pm 4000$~\kms\ originate from the RS just inside the ER. \\\label{fig:rev_shock}}
\end{figure}

Emission from ejecta interacting with the RS at the inner edge of the ER has been observed since the time of the appearance of the first hotspots \citep{Michael1998}. The properties and evolution of the RS have since been studied in detail by many authors (e.g., \citealt{Michael2003,Smith2005,Heng2006,France2010,France2011,Fransson2013,France2015}). The most important aspect for the discussion here is the highest observed velocities, which show that the RS must currently extend well outside the plane of the ER. A recent spectrum of the H$\alpha$ emission from the RS is shown in Figure~\ref{fig:rev_shock}. The spectrum was obtained with the UVES instrument at the Very Large Telescope (VLT) at 11,200 days (see Appendix \ref{app:uves} for details about the observations). The slit covers approximately half of the remnant and the two spectra in Figure~\ref{fig:rev_shock} have been extracted  from the northern and southern halves of the slit. 

The ``horns" at $\sim 4000$~\kms\ are from the RS in the plane of the ER, where the ejecta velocity is $\sim 5500$~\kms\ considering the inclination. However, we also see emission at much higher velocities, which must originate from higher latitudes.  The RS emission from the northern part extends to at least 8500~\kms\ on the blue side, where the narrow [O~I]~$\lambda 6364$ line from the shocked ER interferes. Extrapolating the profile to the background level, we find that the emission extends to $\sim 9000$~\kms.  The red wing  on the southern side is disturbed by the [S II]~$\lambda\lambda 6716, 6731$ doublet. However, there is a faint wing on the red side of these lines above the background, which extends to $\sim 9500$~\kms. The red and blue extensions of the northern and southern parts are consistent with these values.  

Considering that there is some uncertainty in the maximal velocities due to the exact level of the continuum, we conclude that there is evidence of interacting ejecta with velocities at least as high as $\sim 9000$~\kms. Unfortunately, the current observations do not allow us to determine the spatial location of this emission with greater precision than the north/south division presented in Figure~\ref{fig:rev_shock}. We also note that a major part of the new emission outside the ER is outside the slit used for these observations (in particular the south-eastern part, see Appendix~\ref{app:uves}).

\subsubsection{The new spots}
\label{sec:disc-spots}

There are currently about a dozen faint spots located outside the ER of SN~1987A, most of which are in the south-east region. These spots have appeared during the time period between 9500 and 11,500~days (2013 and 2018). In Section~\ref{sec:spots} we presented photometry for the six spots with the highest S/N. Based on the similar fluxes found in the broad R band and the narrow F657N filter we conclude that the emission is due to relatively narrow lines rather than continuum or very broad lines. Only two of the new spots are observed in [O~III]. 

The R-band peak fluxes of the new spots are in the range $1-6 \times 10^{-15}$\ergcms, which is more than an order of magnitude lower than the peak fluxes of the hotspots in the ER. The latter are in the range $4-17 \times 10^{-14}$ \ergcms, with a median of $7 \times 10^{-14}$\ergcms\  (determined from the fits in Section~\ref{sec:expansion}).  The flux decay of most of the new spots is approximately linear with time, just like the ER itself. The decay time scale to reach half the peak flux can be estimated for three of the spots (from the fits in Section~\ref{sec:spots}), giving values of  $\sim 1100, 1600$ and 3200~days. For the spots in the ER the range of time scales is $\sim 2600 - 8000$~days, with a median of $\sim 4300$~days  (see light curves in \citealt{Fransson2015}). 

The emission from the new spots may be due to photoionization by X-rays or shock interaction. We find the former scenario less likely since the time of appearance and the positions of the spots do not correlate with any dramatic increase of the X-ray flux or with the morphology of X-ray emission from the ER \citep{Frank2016}. The rapid appearance and fading of the spots is also hard to explain if they are due to photoionization. In the case of shock interaction, it is possible that the blast wave, which has left the main ER, is now interacting with material further out in the same plane. However, as discussed in Section~\ref{sec:disc-er}, the expected location of the blast wave is consistent with the outer edge of the ER seen in [O~III].  Figures~\ref{fig:rdiff} -- \ref{fig:o3diffs} show that the majority of the new spots are located further out than this. If the spots in the south-east are in the plane of the ER, the blast wave must have propagated significantly faster in this direction, which may be possible given that the ER is fainter in this region. Another possibility is that the spots are due to interaction between fast ejecta and high-latitude material connecting the ER with the ORs. We favor this interpretation because the spatial distribution of all the emission outside the ER can be explained in this way (see Section \ref{sec:disc-diffuse}). We also find that this scenario can account for the emission properties of the spots, as outlined below. 

Explaining the new spots with shock interaction requires that the density of the clumps is high enough and that the shocks propagating into the clumps are slow enough to match the observed fast cooling.  To estimate the shock speed into the clumps we need to estimate the density of the ejecta as a function of velocity. As a rough estimate we assume a spherically symmetric model with a density calculated from the 1D explosion models that have been used to explain the light curve, in particular the ones from  \cite{Shigeyama1990} and \cite{Woosley1988} [also discussed in \citealt{Fransson2013}]. The density profile of the outermost ejecta in the \cite{Shigeyama1990} 14E1 model can be approximated with
 $\rho_{\rm ejecta}(V_{\rm ejecta}) = 7.7\times 10^{-24} (t/10^4 \ {\rm days})^{-3} \ (V_{\rm ejecta}/10^4 \ {\rm km ~s^{-1}})^{-8.6} \ \gccm$, where $t$ is the time since explosion and $V_{\rm ejecta}$ is the velocity of freely expanding ejecta.  The 10H model of \cite{Woosley1988} can be fit with a similar power law, but with the density normalization a factor of 3.0 higher \citep{Michael2003}. The uncertainty in the density is considerable, both from differences in the explosion models and 3D effects. As in \cite{Fransson2013}, we therefore introduce a factor $f$  expected to be in the range $1 \la f \la 3$. 
 We thus assume 
\begin{eqnarray}
\rho_{\rm ejecta}(V_{\rm ejecta}) = 7.7\times 10^{-24}  f  \left({t  \over 10^4 \ {\rm days}}\right)^{-3}\  \nonumber  \\   \left({V_{\rm ejecta}  \over 10^4 \ {\rm km ~s^{-1}}}\right)^{-8.6}   \  \ \gccm.
\label{eq_den}
\end{eqnarray} 

With a number density of the clumps, $n_{\rm H, clump}$, and an He:H abundance of 0.2 by number, the mass density of the clumps are $\rho_{\rm clump} = 3.0 \times 10^{-20} (n_{\rm H,clump}/10^4\ \ccm) \ \gccm$. Estimating the shock velocity in the clumps, $V_{\rm clump}$,  from $\rho_{\rm clump} V_{\rm clump} ^2 \approx \rho_{\rm ejecta} V_{\rm ejecta} ^2 $, where $V_{\rm ejecta}$ is the velocity of freely expanding ejecta at the radial distance of the clumps, we get 
\begin{eqnarray}
V_{\rm clump} \approx 160  f^{1/2}  \left({n_{\rm H, clump}  \over 10^4 \ \ccm }\right)^{-1/2}  \left({V_{\rm ejecta}  \over 10^4 \ {\rm km ~s^{-1}}}\right)^{-3.3} \nonumber  \\  \left({t  \over 10^4  \ {\rm days}}\right)^{-3/2} \ {\rm km ~s^{-1}}. \phantom{aaa}
\label{eq_vcl}
\end{eqnarray}

The temperature behind the shock into the clumps is $T_{\rm e}=1.57 \times 10^5 (V_{\rm clump}/100 \ \rm km ~s^{-1})^2 $~K.
In the coronal approximation, the cooling function from \cite{Dere2009} can be approximated by $\Lambda(T_{\rm e})=2.24 \times 10^{-23} (T_{\rm e}/10^7 \rm\ K)^{-0.75}~{\rm erg\ cm^3\ s^{-1}}$ in the range $10^5 < T_{\rm e} < 2\times 10^7$~K. The cooling time scale, $t_{\rm cool}=3 kT_{\rm e}/\Lambda(T_{\rm e}) n_{\rm e}$, where $n_{\rm e} = 4 n_{\rm H,clump}$ is the electron density behind the shock, can then be expressed as \\

\begin{eqnarray}
t_{\rm cool} \approx 192  f^{1.75}  \left({n_{\rm H, clump}  \over 10^4 \ \ccm }\right)^{-2.75}  \left({V_{\rm ejecta}  \over 10^4 \ {\rm km ~s^{-1}}}\right)^{-11.55}  \nonumber \\  \left( \frac{t}{10^4 \ \rm days} \right)^{-5.25} \rm days \phantom{aaa}
\label{eq:tcool}
\end{eqnarray}
This shows the extreme dependence of the cooling time scale on the ejecta velocity, which is a result of the steep density gradient in the outer part of the ejecta. The fastest ejecta interacting with the RS at 11,200 days have velocities of $\sim 9000$~\kms\ (Section~\ref{sec:disc-rs}), which means that the shock velocity into the clumps will be $\la\ 200 f^{1/2}$~\kms\  if the number density in the clumps is  $\ga\  9 \times10^3\ \ccm$. The corresponding cooling time is $t_{\rm cool} \la\ 478 f^{1.75}$~days, which brackets the range of observed time scales considering that  $1 \la f \la 3$.  The ejecta density close to the RS is only $\sim 8 \times f \ \ccm$.

It is  likely that the highest ejecta velocities at the RS are  $\ga 9000$~\kms \ because of projection effects and the fact that we can only trace fairly bright RS emission above the background continuum. Because of the steep dependence of the cooling time on the ejecta velocity in Equation \ref{eq:tcool}, clumps with densities considerably lower than the above estimate may therefore be able to cool on the observed time scales. Conversely, interaction at  lower ejecta velocities can explain the observations if the clumps have higher densities.  For comparison, we note that clump densities as high as $\ga 10^6\ \ccm$ have been inferred for the winds of some red supergiants and other late-type stars \citep[e.g.,][]{Gray2016}.  

The spectra emitted by the clumps depend on the shock velocities. Shocks with velocities $\ga 200$~\kms\ are expected to give rise to strong [O~III] emission according to the radiative shock models in \cite{Allen2008}. At lower velocities the Balmer lines and  [O II] $\lambda\lambda 3726, 3729$ will dominate. The fact that only two of the dozen new spots have [O~III] emission that is strong enough to be seen in the images is thus indicative of slow shock velocities in most of the clumps. Summarizing, we find it likely that the excitation and fading of the new spots can be explained by shock heating by fast ejecta, although we note that there are considerable uncertainties in both the ejecta structure at high velocities and the density of the clumps.

\subsubsection{The diffuse emission}
\label{sec:disc-diffuse}

\begin{figure*}[t]
\begin{center}
\resizebox{!}{60mm}{\includegraphics{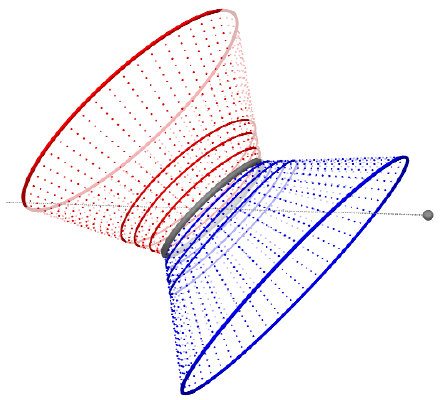}}
\hspace{1.5cm}
\resizebox{!}{60mm}{\includegraphics{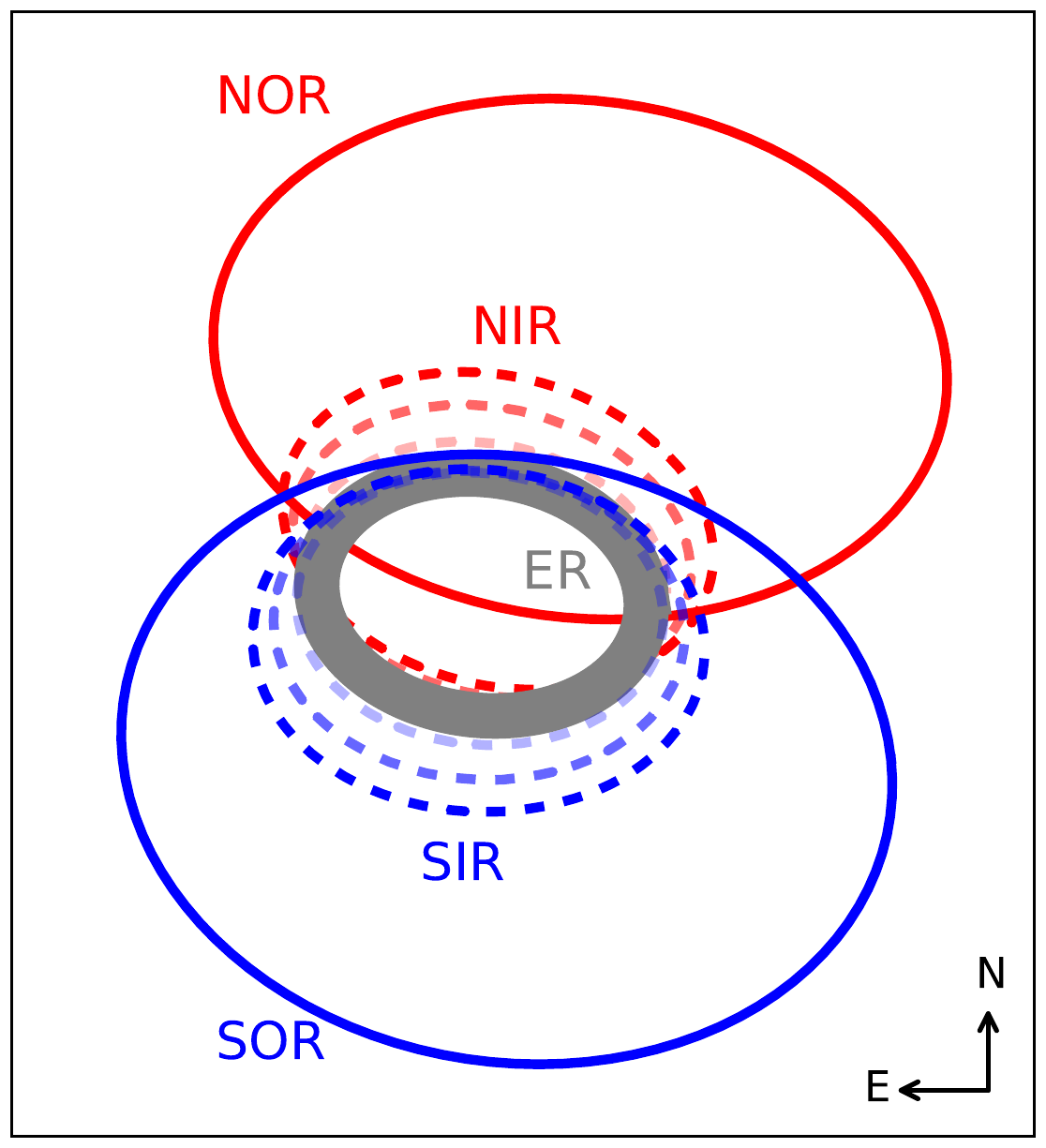}}
\caption{Left: 3D model for the ring nebula surrounding SN~1987A. The ER is shown in grey, the NOR in red and the SOR in blue. The model assumes that there is material connecting the ER and ORs, as illustrated by the red and blue dashed lines. The three additional rings on each side show where spherically symmetric freely expanding ejecta with velocities of 7000, 8000 and 9000~\kms\  would intersect this material at 11,500 days in 2018. These interacting regions make up the intermediate rings (IRs) in the images. The grey sphere and line show the observer and the line of sight, respectively. Right: Projection of the ring system on the sky. The material connecting the rings has been omitted for clarity. The FOV is $5\farcs{0} \times 5\farcs{5}$. \\\label{fig:iringmod}}
\end{center}
\end{figure*}

\begin{figure*}[t]
\plottwo{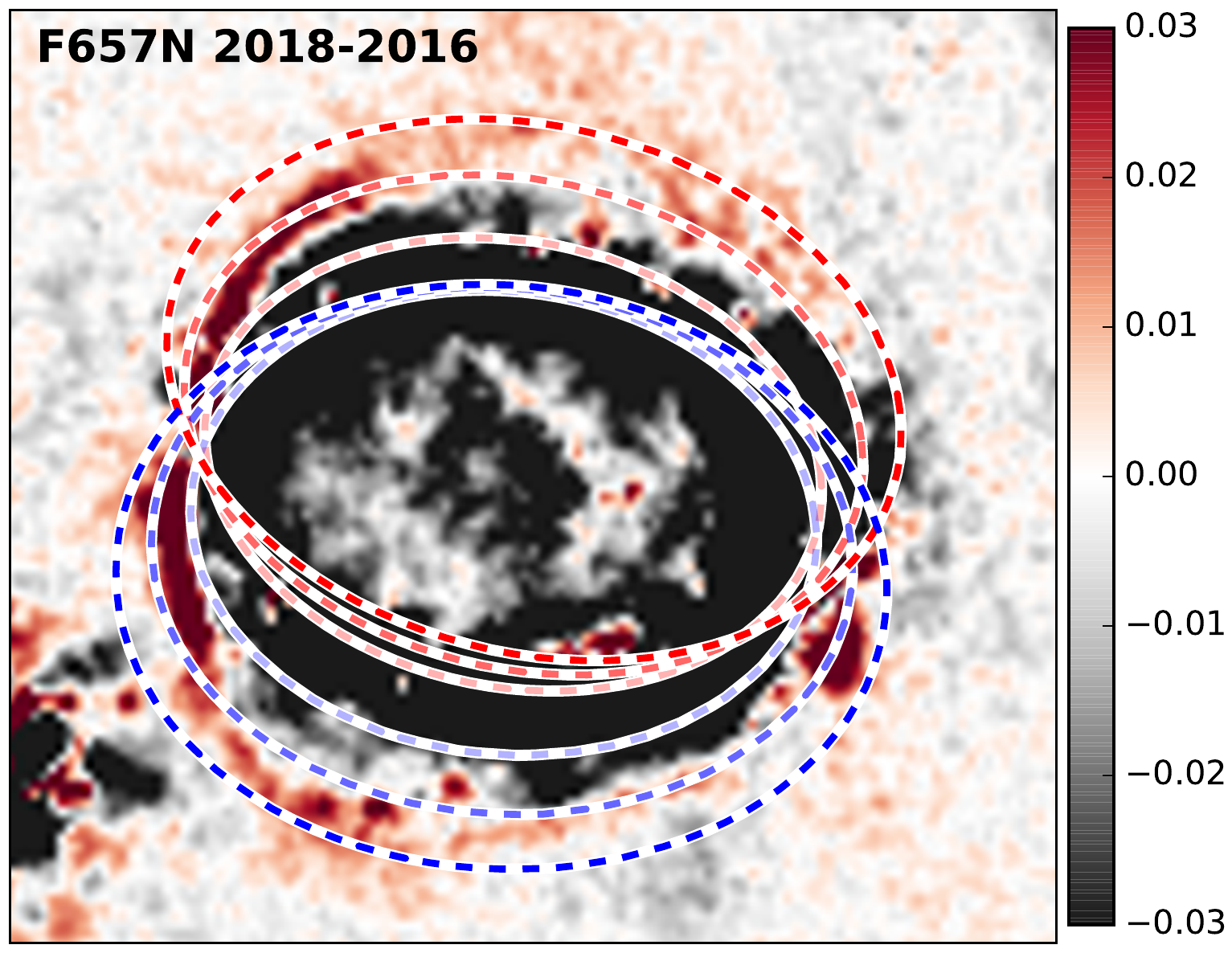}{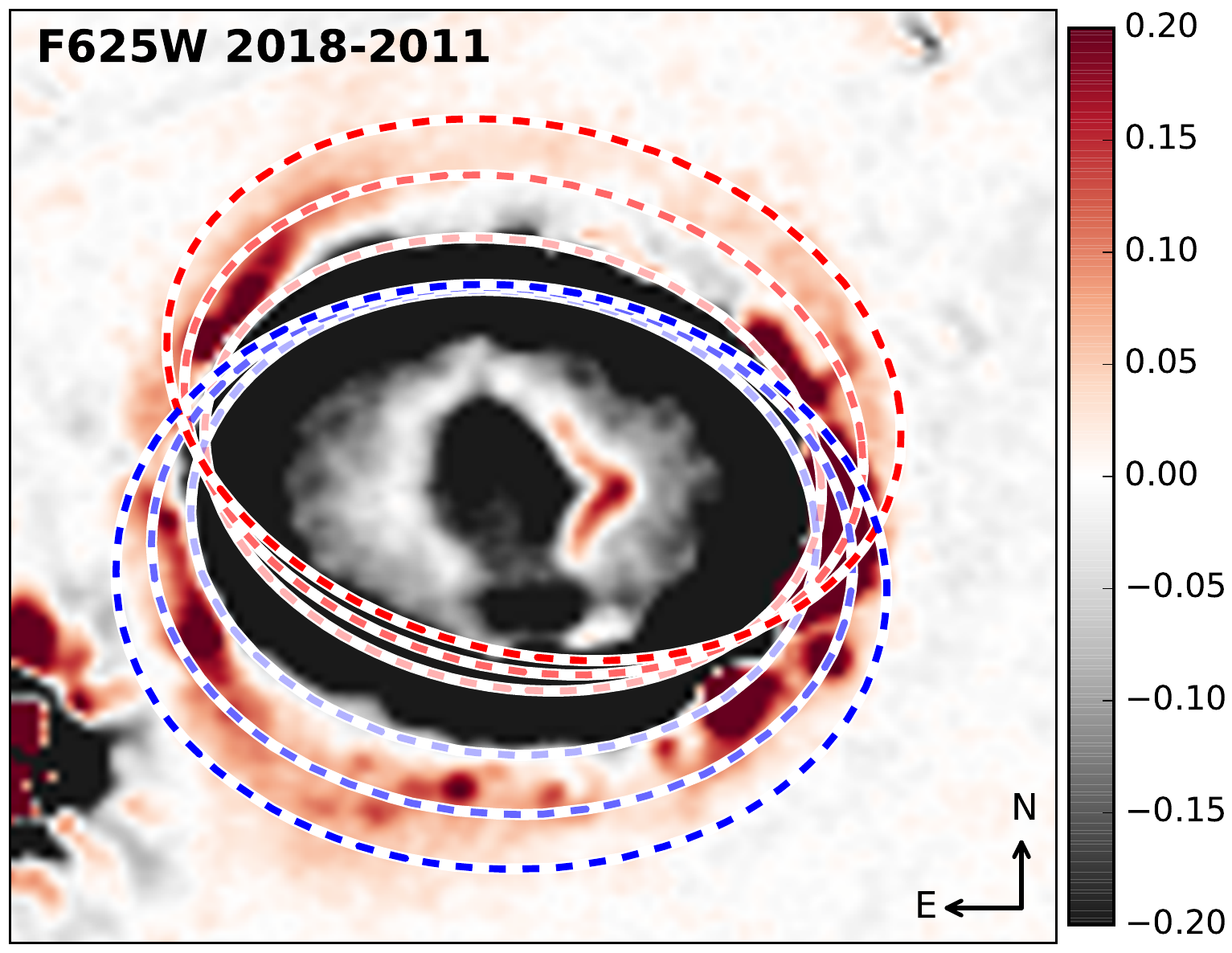}
\caption{Projection of the IRs from the model in Figure~\ref{fig:iringmod},  shown superposed on difference images in H$\alpha$ (left, $11,500-10,700$~days, $2018-2016$) and the R band (right, $11,500-8700$~days, $2018-2011$). The blue rings are on the near side of the ER and the red ones on the far side.  The southern parts of the red rings would thus be obscured by the ER, as shown in Figure~\ref{fig:iringmod}. For each color, the three gradually darker rings correspond to ejecta velocities of 7000, 8000 and 9000~\kms, respectively. The residuals in the south-east corners are due to Star~3 (see Figure~\ref{fig:3ring}). The unit for the color bar is counts~s$^{-1}$ per $0\farcs{025}$ pixel. The FOV is $2\farcs{6} \times 2\farcs{7}$.\\\label{fig:iringproj}}
\end{figure*}

Here we investigate the possibility that the rim of diffuse emission seen in the difference images of H$\alpha$ (Figures~\ref{fig:rdiff} and \ref{fig:hadiffs}) originates in material connecting the ER and ORs. This is motivated by this structure being extended toward the ORs, as well a possible limb brightening on the eastern side. We also note that \cite{France2015} reported some similar structures in narrow-band images obtained at 10,000 days in 2014  and  discussed the possibility of an origin in material connecting the rings. 

To investigate this scenario we set up a simple model for the ring system. We assume that all the rings are circular and that they have inclinations of $43^{\circ}$ (ER), $45^{\circ}$ (NOR) and $38^{\circ}$ (SOR), as reported by \cite{Tziamtzis2011}. The NOR is located on the far side of the ER with respect to the observer and the SOR is on the near side. We take the radius of the ER from our measurements in Section~\ref{sec:expansion} and set the radii of the NOR and SOR to be larger by a factor of 2.2 and 2.3, respectively. The latter were estimated from the latest F657N image in Figure~\ref{fig:3ring}. The distances between the SN and the centers of the NOR and SOR were set to 1.3 and $1.0 \times 10^{18}$~cm, respectively, consistent with the range of distances to the ORs reported by \cite{Tziamtzis2011}. We also apply a small rotation of the system around the north-south axis in order to match the $7^{\circ}$ tilt of the major axis of the ER (obtained from the fits in Section~\ref{sec:expansion}, see also \citealt{Kjaer2010}) and the similar observed tilts of the ORs. Finally, we shift the centers of the NOR and SOR to the west by  5.5  and $4.0 \times 10^{16}$~cm, respectively, which results in a good agreement between the projected model and the images. The fact that there is a small deviation from the symmetry axis has previously been noted in the literature (e.g., \citealt{Burrows1995,Tziamtzis2011}). A 3D view of the model is provided in the left panel of Figure~\ref{fig:iringmod}. 

We assume that there is material connecting the ER with the ORs in the simplest way possible, i.e. straight lines as shown in the left panel of Figure~\ref{fig:iringmod}. The intersection between this material and a spherically symmetric ejecta would result in two intermediate rings (IRs), one on each side. The projection of this system as seen by an observer is shown in the right panel of Figure~\ref{fig:iringmod}. Since the northern part of the southern IR (SIR) and the southern part of the northern IR (NIR) overlap with the ER, the net visible effect is one ring elongated in a similar way as the observed rim of H$\alpha$ emission. The outer boundaries of the IRs in our model correspond to the intersection between freely expanding ejecta traveling at 9000~\kms\ and the material between the rings  at 11,500 days in 2018. This provides an approximate match to the size of the observed structure and is also consistent with the highest velocities in the spectra (Section~\ref{sec:disc-rs}).  In these calculations we have assumed that any deceleration of the ejecta by the H~II region is negligible. The properties of the H~II region at high latitudes are uncertain, but the density likely decreases away from the ER \citep{Chevalier1995,Mattila2010}. 

Figure~\ref{fig:iringproj} shows the projection of the IRs superposed on difference images in F657N ($11,500-10,700$~days, $2018-2016$) and the R band ($11,500-8700$~days, $2018-2011$). The two filters and time scales provide complementary information about the emission. The maximal extent of the diffuse emission is similar in both images, but the R-band image shows brightening in a wider region due to the longer time scale considered. The comparison with the model shows that almost all the emission outside the ER is bracketed by the lines corresponding to 7000 and 9000~\kms. Given the simplicity of the model, the overall agreement is satisfactory. We stress that the purpose of this model is only to show that an origin of the diffuse emission in material connecting the rings is plausible, not to provide an exact fit to the observations. More generally, IRs above and below the plane of the ER  would result from the intersection between the ejecta and any other bipolar, axisymmetric structure, such as an hourglass \citep{Chevalier1995,Crotts1995}. We also stress that the IRs are not well-defined rings in the same way as the ER and ORs and that they only clearly appear as rings in the difference images. The IRs are seen to expand with time, but the S/N of the diffuse emission in difference images  between consecutive years is too low to determine the expansion velocity (see Figure~\ref{fig:rdiff}). 

Assuming that the diffuse emission originates from ejecta reaching the RS we expect strong H$\alpha$ emission and no [O~III], which agrees with the observations. In order to further test the model we would need spectral information to obtain Doppler shifts for the different parts of the diffuse emission. Under the assumption that the emitting material has the same velocity as the freely expanding ejecta, both IRs will be blueshifted in the north and redshifted in the south. Our model with ejecta velocity  9000~\kms\ (see Figure~\ref{fig:iringmod}), gives a maximum blueshift in the north of the NIR of $\sim -3400$~\kms\ and a maximum redshift in the south of $\sim 8400$~\kms. The corresponding numbers for the SIR are $\sim -7700$~\kms\ in the north and  $\sim 3700$~\kms\  in the south.  This means that some of the emission from the IRs should be shifted outside the F657N filter, which covers $[-3500,+3600]$~\kms\ around H$\alpha$ at the systematic velocity of SN~1987A. However, the main parts of the IRs that are shifted outside the filter also overlap with the ER, which means that there will not be a noticeable effect in the images. 

Basic velocity information can instead be obtained from previous observations in other narrow filters, as discussed in \cite{France2015}. The F645N filter probes blueshifted H$\alpha$ in the range $[-7900, -2700]$~\kms\ (Table~\ref{tab:lines}). The F645N difference image between 10,700 and 10,000~days ($2016-2014$) shows a rim of emission that follows the trends expected for the SIR (Figure~\ref{fig:hadiffs}, right panel). In particular, there is no emission in the southern part of the ring, which is expected to be redshifted outside the filter. The fact that the NIR is not seen in F645N is also consistent with the expectations from the model since almost all the emission from NIR is expected at velocities redshifted outside this filter. On a detailed level, the shape of the SIR in F645N is somewhat different from that seen in F657N, which may be explained by the different epochs and the different velocities probed. The SIR also disappears outside the filter further to the south than expected from our model, suggesting interaction at higher velocities and/or that the material connecting the rings may have a more parabolic shape with respect to the polar axis. Such a shape would imply interaction closer to the line of sight and thus higher blueshifts. Emission from the RS close to the line of sight has also been inferred from previous two-dimensional spectra of the RS obtained with {\it HST}/STIS (e.g., \citealt{France2010,Fransson2013}).

To probe redshifted H$\alpha$ emission we can use the F665N filter, which covers the interval [300,7600]~\kms. There has only been one observation in this filter (at 10,000 days in 2014), so we cannot investigate difference images as above. However, \cite{France2015} noted streaks of emission in this image, similar to the expectations for the NIR. These features are especially clear when subtracting the F658N filter, which removes much of the emission from the ER itself.  The likely  emission from the NIR seen in F665N also includes weak emission north of the ER. Due to the filter limits, this cannot be the blueshifted H$\alpha$ predicted by the model. As discussed for the SIR above, this may suggest a more parabolic geometry for the high-latitude material. In summary, we conclude that the approximate velocity information obtained from the narrow filters is consistent with the trends expected from the model, but that a more complex model will be needed to fully account for the observations.

With spatially resolved spectral information from future observations, it will be possible to obtain information about the detailed geometry and density of the high-latitude material. This information will be important for discriminating between different models for the formation of the triple-ring nebula. For example, the binary merger model by \cite{Morris2009} does not predict material connecting the rings, whereas the scenario proposed by \cite{Smith2013} predicts high-latitude material in a parabolic shape. We also stress that the current observations only probe high-latitude material extending $\sim 25\%$ of the distance toward the ORs. Future monitoring will be required to determine if there is material all the way out to the ORs. 
 
The fact that there may be material between the ER and the ORs has previously been suggested by light echoes, with \cite{Crotts1995} inferring an hour-glass structure and \cite{Sugerman2005} a more cylindrical shape. In addition, in an analysis of early {\it HST} images, \cite{Burrows1995} noted that the ER and ORs may be connected by low-surface brightness walls that fade with distance from the ER. The possibility of significant diffuse emission inside the rings in {\it HST} images was also noted by \cite{Tziamtzis2011}. Such diffuse emission is expected due to photoionization of the material between the rings by the initial SN flash (see \citealt{Lundqvist1996} for a model) and later on by the X-ray emission from the ER. The latter most likely has a subdominant effect at current epochs, given the lack of temporal and spatial correlation between the diffuse emission and the X-ray emission from the ER, as well as the fact that the shock-interaction model discussed above can explain the main properties of the emission. The long-term evolution of the diffuse emission and the effect of X-rays will be studied in a separate paper.

\subsection{Evolution of the ejecta}
\label{sec:disc-ejecta}

The flux of the inner ejecta in SN 1987A  has evolved significantly with time, as seen in Sections \ref{sec:images} and \ref{sec:lightcurves}. From our previous study of the ejecta light curve \citep{Larsson2011}, we found that the ejecta faded as expected from radioactive decays until $\sim 5000$~days. After this point the flux started to increase, which can be explained by energy input from the X-ray emission from the ER.  Here we have extended the light curve until 11,500 days and also investigated the eastern and western parts separately. In order to follow the freely expanding ejecta, we adjusted the size of the aperture to always correspond to a velocity of 2800\ \kms\ in the plane of the sky. We note that this means that the southern part of the ejecta, which overlaps with the ER in the recent images, is outside the aperture (see Figure~\ref{fig:apertures}). This region is significantly affected by X-rays (see \citealt{Larsson2013}), which means that the fraction of  flux from the full ejecta that falls outside the aperture is expected to increase with time. 

We find that the two halves of the ejecta evolve  similarly during the decaying phase and initial rise, brightening by a factor of 2.6 in the R band between the minimum at 5400 days and 7200~days.  The increase continues at a slower rate thereafter, especially in the eastern half, where the flux reached a constant level at 10,700 days.  The western half makes up 57\% of the total R-band flux in the last observation at 11,500~days.  The B band evolves in the same way, but with a somewhat smaller flux increase.      

These trends are in general agreement with the expectations from powering of the ejecta by X-rays. The soft (0.5-2~keV) X-ray light curve has leveled off in recent years, reaching a constant level at 9500 days \citep{Frank2016}. In addition, the X-ray emission from the western side of the ER started to dominate  between 7000 and 8000 days, making up just over 60\% of the total X-ray flux at $10,500$~days \citep{Frank2016}. While this is similar to the evolution of the ejecta, we note that the soft X-ray light curve does not directly correlate with the ejecta light curve. This is not surprising given the complexity of the energy deposition, which depends on the X-ray flux at different energies (including very soft X-rays below the observed limit), as well as geometric effects, with the asymmetric ejecta being illuminated by a non-uniform ring in the equatorial plane. In addition, the aperture used for the flux measurements does not include the ejecta closest to the ring, where X-ray emission below  $\sim 1$~keV is expected to be absorbed \citep{Fransson2013}.  

An edge-brightened morphology of the ejecta is expected since most of the X-ray emission will be absorbed in the outer ejecta, as discussed in detail in \cite{Fransson2013} and \cite{Larsson2013}. This is supported by observations since the morphology becomes edge-brightened at the same time as the flux starts increasing (see Figures~\ref{fig:allred} and \ref{fig:allblue}). In the most recent images the morphology is dominated by a bright clump in the western ejecta, as expected from the fact that the X-ray emission from the ER is stronger on the western side \citep{Frank2016}. The brightening of this clump has also significantly affected the H$\alpha$ line profile of the ejecta, as shown in \cite{Larsson2016}. The strong impact on the H$\alpha$ line is also reflected in the larger flux increase in the R band compared to the B band. The lines in the B band (including Fe~I lines from the inner ejecta, see Table~\ref{tab:lines}) are generally expected to be less affected by the X-rays, although a detailed modeling would be required to quantify the differences.

\subsection{Future evolution of SN 1987A}
\label{sec:disc-future}

From the recent {\it HST} observations studied in this paper we have found that the emission from the ER is fading linearly with time, that more emission is appearing outside the ER and that the flux increase of the ejecta has leveled off. We have also seen that significant changes take place on a one-year time scale. Future observations are expected to show a continued strong evolution of the remnant. An extrapolation of the R-band light curve of the ER shows that the flux will reach zero by  $\sim 2035$. This is a very rough prediction of when the optical ER will be completely gone, considering that the rate of the flux decay may change with time and that there are large difference in flux between different parts of the ER. The south-east part, which is the faintest, will most likely have disappeared well before 2035 (Figure~\ref{fig:ringlc}, right panel). 

At the same time as the ER is fading we expect the optical emission from high-latitude material outside the ER to become increasingly dominant.  Studies of this emission will shed new light on the formation of the rings and the mass-loss history of the progenitor star. In particular, the faint spots outside the ER will provide information about clumping in the wind(s) of the progenitor. Additional spots may appear with time and  their densities can be constrained from spectral observations. From spectra of the regions outside the ER it will also be possible to determine if the diffuse H$\alpha$ emission comes from freely expanding ejecta interacting with a RS, and, assuming this is the correct interpretation,  determine the 3D structure of the high-latitude material. We also expect the RS, traced by the IRs in the difference images, to expand with time. Future {\it HST} imaging observations may allow us to constrain the expansion speed, which in turn places constraints on the density of the material between the clumps. 

The interaction outside the ER may also produce emission at other wavelengths, including radio and X-rays, depending on the densities involved. The prospects for detecting multiwavelength emission from the optically emitting regions outside the ER depend on the relative fluxes of the ER and the high-latitude material at those wavelengths, as well as the spatial resolution. The brightest optical emission outside the ER is still more than an order of magnitude fainter than the ER itself, and we are only able to identify it owing to the excellent spatial resolution of {\it HST}. 

On a time scale of decades to centuries, we expect to observe interaction between the ejecta and the complex circumstellar environment outside the triple-ring nebula. This environment comprises several distinct structures, including the so-called Napoleon's hat (see \citealt{Sugerman2005} for details). We expect the ejecta to start interacting with the innermost of these structures on a time-scale of $\sim 20$~years, while it will take $\sim  300$~years before ejecta traveling at $~10,000$~\kms\ reach the outermost parts of Napoleon's hat (assuming the dimensions from \citealt{Sugerman2005}). 

The evolution of the ejecta is strongly coupled to the evolution of the X-ray emission from the ER. The flux from the ejecta has already started to level off, responding to the evolution of the X-ray light curve. If the X-ray emission from the ring starts fading we expect this to be reflected in the ejecta. However, as the ejecta expand we also expect the X-rays to penetrate to lower velocities. This  is illustrated by the models in  \cite{Fransson2013}, which show that X-rays with energies above $\sim 2$~keV penetrate into the metal-rich core at 11,000~days, while only X-rays above 5~keV (where the flux is very low) reaches the core at 7000~days. Although the exact evolution depends on the simplified 1D structure of the ejecta used in these simulations, the general trend should be robust. We therefore expect the X-rays to start affecting the core in the future, resulting in changes of the metal lines (e.g.,  [Fe II] + [Si II] at 1.644~$\mu$m) as well as the molecular emission (e.g., H$_2$, CO and SiO). 

Finally, we expect a dramatic evolution as ejecta with increasingly high densities reach the ER. The southern part of the ejecta already overlaps with the ER in the images, but most of this material is well above the plane of the ring, moving closer to the plane of the sky \citep{Larsson2016}. There is instead more ejecta in the plane of the ring in the north. The velocity of the ejecta that are just reaching the RS in the plane of the ER is currently $\sim  5500$~\kms. It is highly likely that we will start seeing strong signs of interaction within the next five years when this velocity has decreased to $4500$~\kms, which approximately corresponds to the outer edges of the inner ejecta seen in the {\it HST} images.  \\

\section{Conclusions}
\label{sec:conclusions}

In this paper we have presented an analysis of the recent evolution of SN~1987A as probed by {\it HST} imaging observations. Our main results are summarized below.

\begin{itemize}

\item The optical emission from the ER continues to fade linearly with time in all filters,  although the light curve in the F502N~[O III] filter peaks earlier and decays faster than the R and B bands by $\sim~1000$~days. This can be explained by the fact that the [O~III]  emission reflects the instantaneous energy input from the shocks, while the low-ionization lines in the other filters reflect the accumulated mass behind the shocks. 

\item The ER is expanding at $680 \pm 50$~\kms. This number is independent of the inclination and corresponds to the typical velocity of transmitted shocks in the dense clumps/hotspots. The current radius of the ER determined from the hotspots is $0\farcs{82}$. 

\item The image of the ER in [O~III] exhibits a kind of double structure, with an inner part matching the observations in other filters and a fainter diffuse component with a  semi-major axis of $\sim 1\farcs{0}$. The outer structure is consistent with being due to material swept up by the blast wave propagating in low-density gas between the clumps. Alternatively, the double structure may have formed by ejection episodes at different times and/or different velocities. 

\item A dozen new spots have appeared outside the ER since 9500~days. Most of these are located in the south-east. Compared to the hotspots in the ER, the new spots are significantly fainter (by one to two orders of magnitude at peak flux) and fade faster with time. 

\item A rim of diffuse H$\alpha$ emission has appeared outside the ER. This component is very prominent in the last observation from 11,500~days, but first started to emerge around the    same time as the new spots. The rim is extended toward the ORs rather than having the same shape as the ER.  

\item The new spots and diffuse emission outside the ER can be explained by fast ejecta ($\sim 9000$~\kms) interacting with high-latitude material that extends from the ER toward the ORs. In this scenario the spots are due to slow shocks ($\sim 200$~\kms) driven into dense clumps, while the diffuse emission is due to fast ejecta interacting with a RS. Observations providing spatially resolved spectral information are needed to further test this scenario and determine the precise geometry of the high-latitude material. 

\item The ejecta continue to brighten, but at a gradually slower rate.  The eastern and western halves evolved similarly until $7000$~days, after which the western part has brightened faster and remained brighter. These trends are in agreement with the scenario that X-ray emission from the ring is powering the optical emission from the ejecta. 

\end{itemize}

\acknowledgments
The authors would like to acknowledge the contributions of the deceased Arlin Crotts to understanding the circumstellar environment of SN 1987A. This work was supported by the Knut and Alice Wallenberg Foundation and the Swedish Research Council.  RAC acknowledges support from NSF grant 1814910. The research of JCW is supported by NSF AST-1813825. Support for HST GO program numbers 13810, 14333, 14753 and 15256  was provided by NASA through grants from the Space Telescope Science Institute, which is operated by the Association of Universities for Research in Astronomy, Inc., under NASA contract NAS5-26555. The ground-based observations were collected at the European Organization for Astronomical Research in the Southern Hemisphere, Chile (ESO Program 100.D-0-705(A)).

%

\vspace{5mm}
\facilities{HST (WFC3, ACS, WFPC2), VLT (UVES)}



{\software{DAOPHOT \citep{Stetson1987}, 
	DrizzlePac \citep{Gonzaga2012},
	matplotlib \citep{Hunter2007},
	Mayavi \citep{Ramachandran2011}}



\appendix

\section{Light curve tables}
\label{app:lc-tables}

Here we provide the flux measurements for the ER and the ejecta in Tables~\ref{tab:ringlc} and \ref{tab:ejectalc}, respectively. The apertures are shown in Figure~\ref{fig:apertures} and the fluxes are plotted in the left panels of Figures~\ref{fig:ringlc} and \ref{fig:ejectalc}. 

\begin{deluxetable}{llcccc}[t]
\tablecaption{Flux evolution of the ER  \label{tab:ringlc}}
\tablecolumns{5}
\tablenum{5}
\tablewidth{0pt}
\tablehead{
\colhead{Date} & 
\colhead{Epoch\tablenotemark{a}} & 
\colhead{R-band flux} &
\colhead{B-band flux} & 
\colhead{F502N flux} & \\
\colhead{(YYYY-mm-dd)} &
\colhead{(d)} &
\colhead{($10^{-13}$~\ergcms)} &
\colhead{($10^{-13}$~\ergcms}) &
\colhead{($10^{-14}$~\ergcms}) & 
}
\startdata
1994-02-03 & 2537 & \nodata & \nodata    	&  $4.16 \pm 0.06$ 	\\
1994-09-24 & 2770 & $6.111 \pm 0.007$	        & $1.059 \pm 0.006$		& \nodata	\\
1995-03-05 & 2932 & $6.047 \pm 0.007$		& $0.995 \pm 0.006$		& \nodata	\\
1996-02-06 & 3270 & $5.775 \pm 0.007$		& $1.010 \pm 0.005$	 	& $3.19 \pm 0.04$	\\
1997-07-10 & 3790 & $5.250 \pm 0.007$		& $0.943  \pm 0.006$	&  \nodata	 \\
1997-07-12 & 3792 & \nodata   & \nodata  			&  $2.63 \pm 0.04$ 	\\
1998-02-06 & 4001 & $5.073 \pm 0.008$		& $1.034 \pm 0.007$		&  \nodata	 \\
1999-01-07 & 4336 & $4.766 \pm 0.004$		& $1.106 \pm 0.005$		&  \nodata	 \\
1999-04-21 & 4440 & $4.837 \pm 0.008$		& $1.178 \pm 0.006$		&  \nodata	 \\
2000-02-02 & 4727 & $4.779 \pm 0.008$		& $1.081 \pm 0.008$		&  \nodata	 \\
2000-06-16 & 4862 & $4.750 \pm 0.008$		& $1.071 \pm 0.006$		&   $2.59 \pm 0.06$ \\
2000-11-13 & 5012 & $4.926 \pm 0.003$		& $1.165 \pm 0.005$		& \nodata	\\
2000-11-14 & 5013 & \nodata   & \nodata  			& $ 2.80 \pm 0.05$ 	\\
2001-03-23 & 5142 &  $5.119 \pm 0.007$		& $1.277 \pm 0.006$		& \nodata	\\
2001-12-07 & 5401 & $5.578 \pm 0.006$		& $1.484 \pm 0.005$		& $ 3.16 \pm 0.05$ \\
2003-01-05 & 5795 & $8.137 \pm 0.004$		&  $2.345 \pm 0.002$	&  $ 6.88 \pm 0.04$ 	\\
2003-08-12 & 6014 & $9.760 \pm 0.005$		& $2.935 \pm 0.003$		& \nodata	\\
2003-11-28 & 6122 &  $10.523 \pm 0.004$		& $3.259 \pm 0.002$		& $ 9.34 \pm 0.07$ 	\\
2004-12-15 & 6505 & \nodata	& $4.381 \pm 0.002$		& $11.41  \pm 0.08$	\\
2005-04-02  & 6613 &  \nodata	& $4.709 \pm 0.003$		&  \nodata	\\
2005-09-26 & 6790 & $15.762 \pm 0.001$	& \nodata				&  \nodata	 \\
2005-11-18 & 6843 & \nodata   & \nodata  			&  $ 12.79 \pm 0.08$	\\
2006-04-15 & 6991 & $17.406 \pm 0.004$	&  $5.906 \pm 0.003$	&  \nodata	 \\
2006-12-08 & 7228 & \nodata   & \nodata  			&  $ 13.11 \pm 0.10$\\
2006-12-06 & 7226 & $18.645 \pm 0.004$	& $6.418 \pm 0.003$ 	& \nodata	\\
2007-05-12\tablenotemark{b} & 7383 & $19.040 \pm 0.006$	& $6.795 \pm 0.008$ & \nodata	\\
2008-02-19\tablenotemark{b} & 7666 & $19.997 \pm 0.007$	& $7.311 \pm 0.009$	 & \nodata	\\
2009-04-29\tablenotemark{b} & 8101 & $20.898	 \pm 0.008$& $7.774 \pm 0.010$	& \nodata	\\
2009-12-12 & 8328  &  $20.876 \pm 0.003$	& $7.870 \pm 0.005$ 	& $12.38 \pm 0.10$ \\
2011-01-05 & 8717 & $20.257 \pm 0.006$	& $7.931 \pm 0.004$	 	& \nodata	\\
2013-02-06 & 9480 & $18.416 \pm 0.005$	& $7.392 \pm 0.004$		& \nodata	\\
2014-06-15 & 9974 & $17.234 \pm 0.005$	& $6.973 \pm 0.004$		& \nodata	\\
2014-06-20 & 9979 &  \nodata  & \nodata  			& $ 8.29	\pm 0.06$ \\
2015-05-24 & 10,317 & $16.876 \pm 0.005$  &  $6.796  \pm 0.004$	& \nodata	 \\
2016-06-08 & 10,698 & $16.032 \pm 0.006$  &  $6.453 \pm 0.006$	  	& $ 7.26 \pm 0.09$ \\
2017-08-03 & 11,119 &  $14.783 \pm 0.004$  &  $5.986 \pm 0.004$ 	&  $ 6.53 \pm 0.07$ \\
2018-07-08 &  11,458 & $14.002  \pm 0.004$ &  $5.673 \pm 0.004$ 	& \nodata	 \\
2018-07-10& 11,460  & \nodata & \nodata 			& $ 5.86 \pm 0.05$	\\
\enddata
\tablecomments{All fluxes have been normalized to the bandpass of the WFPC2 filters as described in Section~\ref{sec:obs}. The error bars on fluxes are statistical only. We refer the reader to Sections~\ref{sec:obs} and \ref{sec:lightcurves} for information about the systematic uncertainties.}
\tablenotetext{a}{Days since explosion on 1987-02-23.}
\tablenotetext{b}{Fluxes are uncertain due to significant CTE loss in WFPC2, see Sections~\ref{sec:obs} and \ref{sec:lightcurves} for details. }
\end{deluxetable}

\begin{deluxetable}{llcccc}[t]
\tablecaption{Flux evolution of the ejecta  \label{tab:ejectalc}}
\tablecolumns{4}
\tablenum{6}
\tablewidth{0pt}
\tablehead{
\colhead{Date} & 
\colhead{Epoch\tablenotemark{a}} & 
\colhead{R-band flux} &
\colhead{B-band flux} & \\
\colhead{(YYYY-mm-dd)} &
\colhead{(d)} &
\colhead{($10^{-14}$~\ergcms)} &
\colhead{($10^{-14}$~\ergcms}) &
}
\startdata
1994-09-24 & 2770 & $3.34 \pm 0.02$ & $2.65 \pm 0.03$	\\
1995-03-05 & 2932 & $3.10 \pm 0.02$ & $2.51 \pm 0.03$  \\
1996-02-06 & 3270 & $2.62 \pm 0.01$ & $2.05\pm 0.02$	 \\
1997-07-10 & 3790 & $2.05  \pm 0.01$ & $1.66 \pm 0.02$ \\
1998-02-06 & 4001 & $1.94 \pm 0.02$  & $1.60 \pm 0.03$ \\
1999-01-07 & 4336 & $1.763 \pm 0.008$ & $1.39 \pm 0.02$  \\
1999-04-21 & 4440 & $1.70 \pm 0.01$ & $1.34 \pm 0.02$  \\
2000-02-02 & 4727 & $1.69 \pm 0.01$ & $1.24 \pm 0.03$ \\
2000-06-16 & 4862 & $1.62 \pm 0.01$ & $1.23 \pm 0.02$ \\ 
2000-11-13 & 5012 &  $1.610 \pm 0.006$  & $1.21 \pm 0.02$ \\
2001-03-23 & 5142 & $1.67 \pm 0.01$  & $1.24 \pm 0.02$ \\
2001-12-07 & 5401 & $1.58 \pm 0.01$  & $1.16 \pm 0.01$ \\
2003-01-05 & 5795 & $2.077\pm 0.006$ & $1.469 \pm 0.005$ \\
2003-08-12 & 6014 & $2.259\pm 0.008$ & $1.483 \pm 0.006$ \\
2003-11-28 & 6122 & $2.286 \pm 0.006$ & $1.533 \pm 0.004$ \\
2004-12-15 & 6505 & \nodata	& $1.864 \pm 0.005$\\
2005-04-02  & 6613 & \nodata	& $1.927 \pm 0.006$ \\
2005-09-26 & 6790 & $3.311 \pm 0.002$ & \nodata \\
2006-04-15 & 6991 &  $3.684 \pm 0.006$ &	$2.164 \pm 0.006$ \\
2006-12-06 & 7226 & $4.083\pm 0.007$ & $2.431 \pm 0.005$ \\
2007-05-12\tablenotemark{b} & 7383 & $4.177\pm 0.009$ & $2.39\pm 0.01$ \\
2008-02-19\tablenotemark{b} & 7666 & $4.54 \pm 0.01$ & $2.41 \pm 0.02$ \\
2009-04-29\tablenotemark{b} & 8101 & $5.23 \pm 0.01$ & $2.81 \pm 0.02$ \\
2009-12-12 & 8328  & $5.370 \pm 0.005$ & $2.74 \pm 0.01$ \\
2011-01-05 & 8717 & $5.77 \pm 0.01$  & $2.909 \pm 0.008$ \\
2013-02-06 & 9480 & $6.201\pm 0.009$  & $3.019 \pm 0.008$ \\
2014-06-15 & 9974 & $6.443 \pm 0.009$  & $3.155 \pm 0.009$ \\
2015-05-24 & 10,317 & $6.720 \pm 0.009$ & $3.239 \pm 0.009$ \\
2016-06-08 & 10,698 & $7.02 \pm 0.01$  & $3.40 \pm 0.01$ \\
2017-08-03 & 11,119 &  $7.07  \pm 0.01$ & $3.348 \pm 0.008$ \\
2018-07-08 &  11,458 & $7.18 \pm 0.01$  & $3.449 \pm 0.009$ \\
\enddata
\tablecomments{All fluxes have been normalized to the bandpass of the WFPC2 filters as described in Section~\ref{sec:obs}. The error bars on fluxes are statistical only. We refer the reader to Sections \ref{sec:obs}, \ref{sec:lightcurves} and Appendix \ref{app:tinytim} for information about the systematic uncertainties.}
\tablenotetext{a}{Days since explosion on 1987-02-23.}
\tablenotetext{b}{Fluxes are uncertain due to significant CTE loss in WFPC2, see Sections~\ref{sec:obs} and \ref{sec:lightcurves} for details. }
\end{deluxetable}

\clearpage

\section{Models for the ER}
\label{app:tinytim}

\begin{figure}[t]
\begin{center}
\resizebox{50mm}{!}{\includegraphics{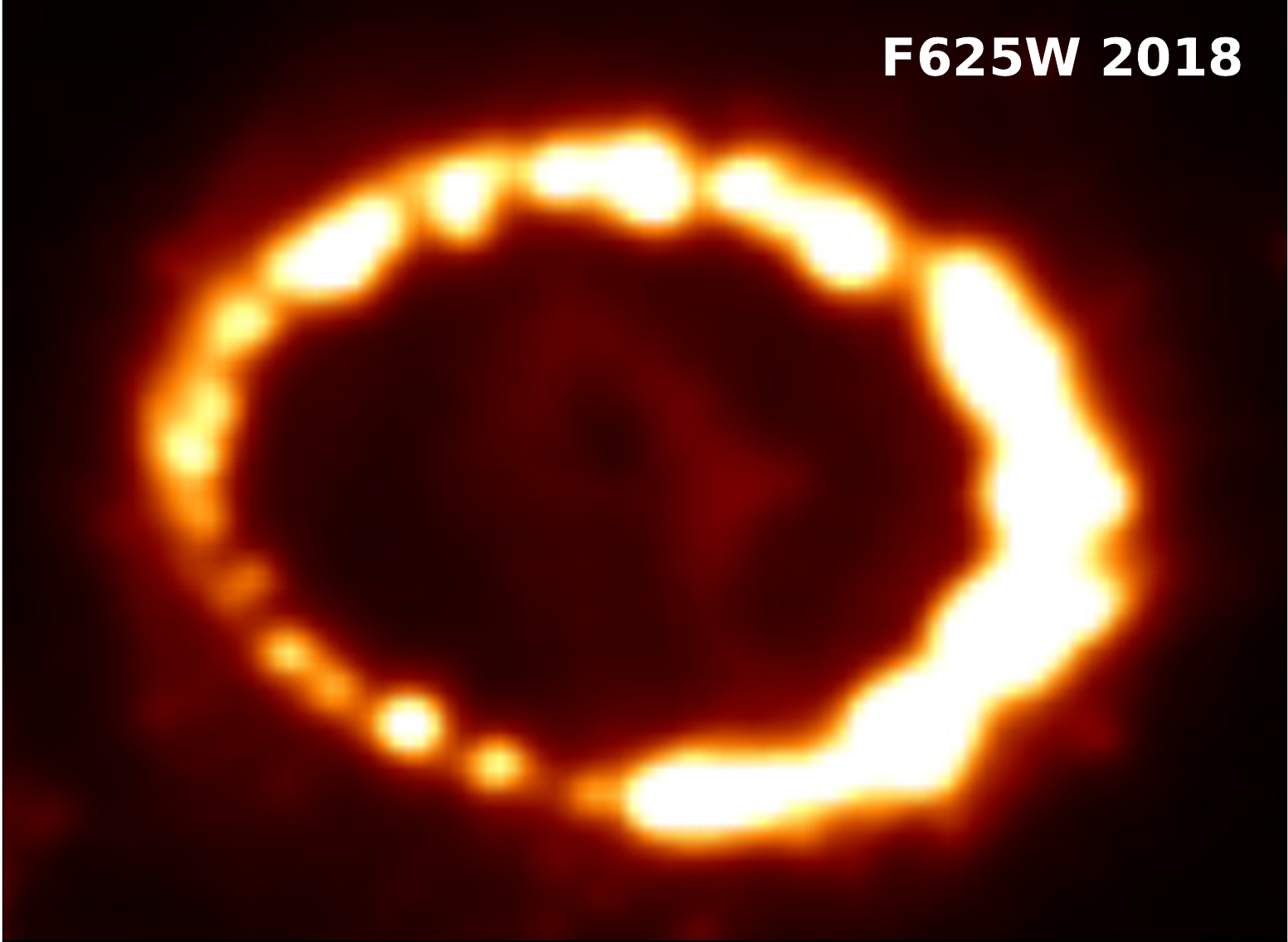}}
\resizebox{50mm}{!}{\includegraphics{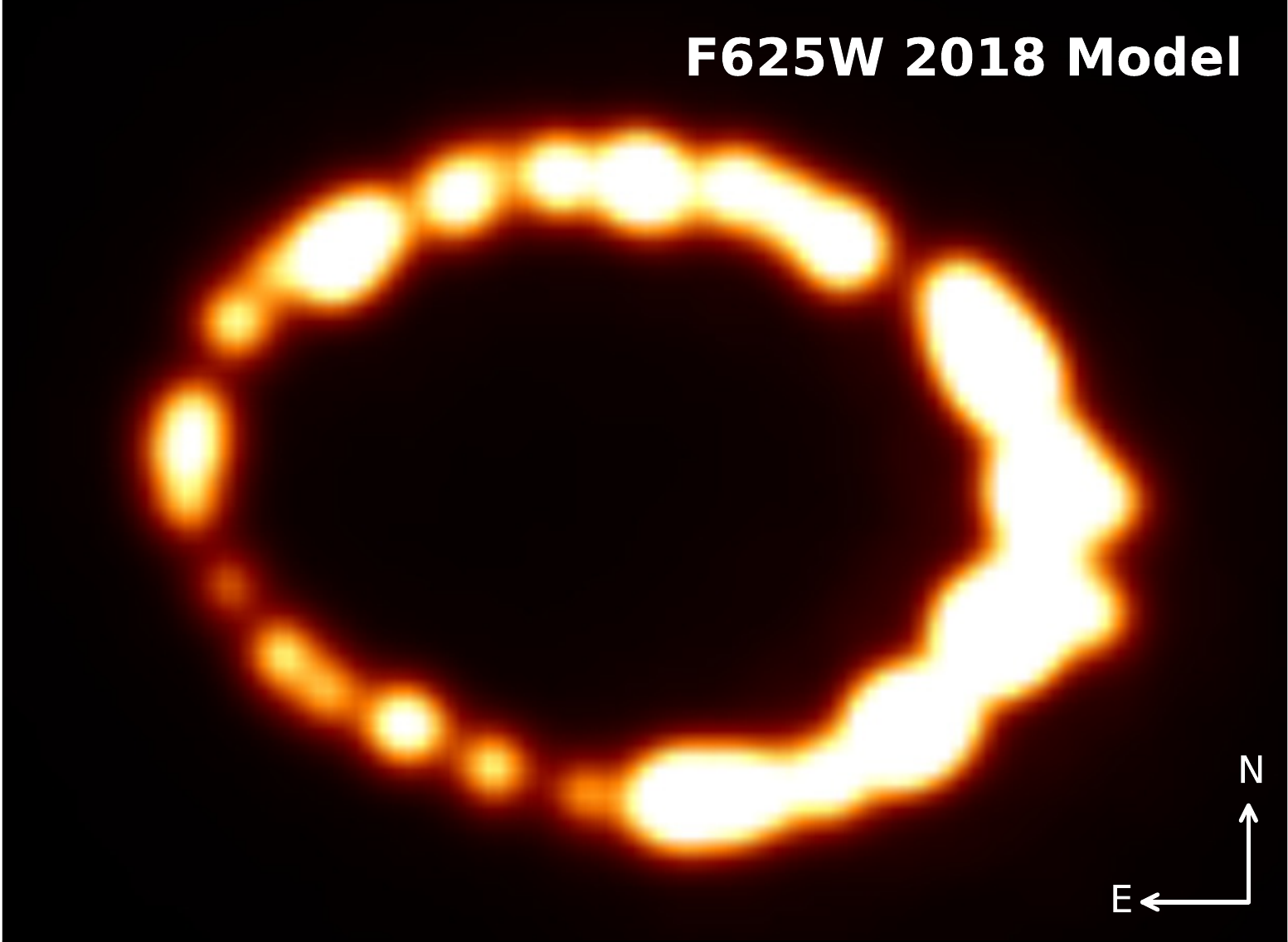}}
\caption{R-band image from 11,500 days (2018) together with the corresponding model for the ER (left and right panels, respectively). The image scale is linear. The FOV is $2\farcs{5} \times 1\farcs{8}$. \\ 
\label{fig:tinymodel}}
\end{center}
\end{figure}
\begin{figure}[t]
\begin{center}
\resizebox{70mm}{!}{\includegraphics{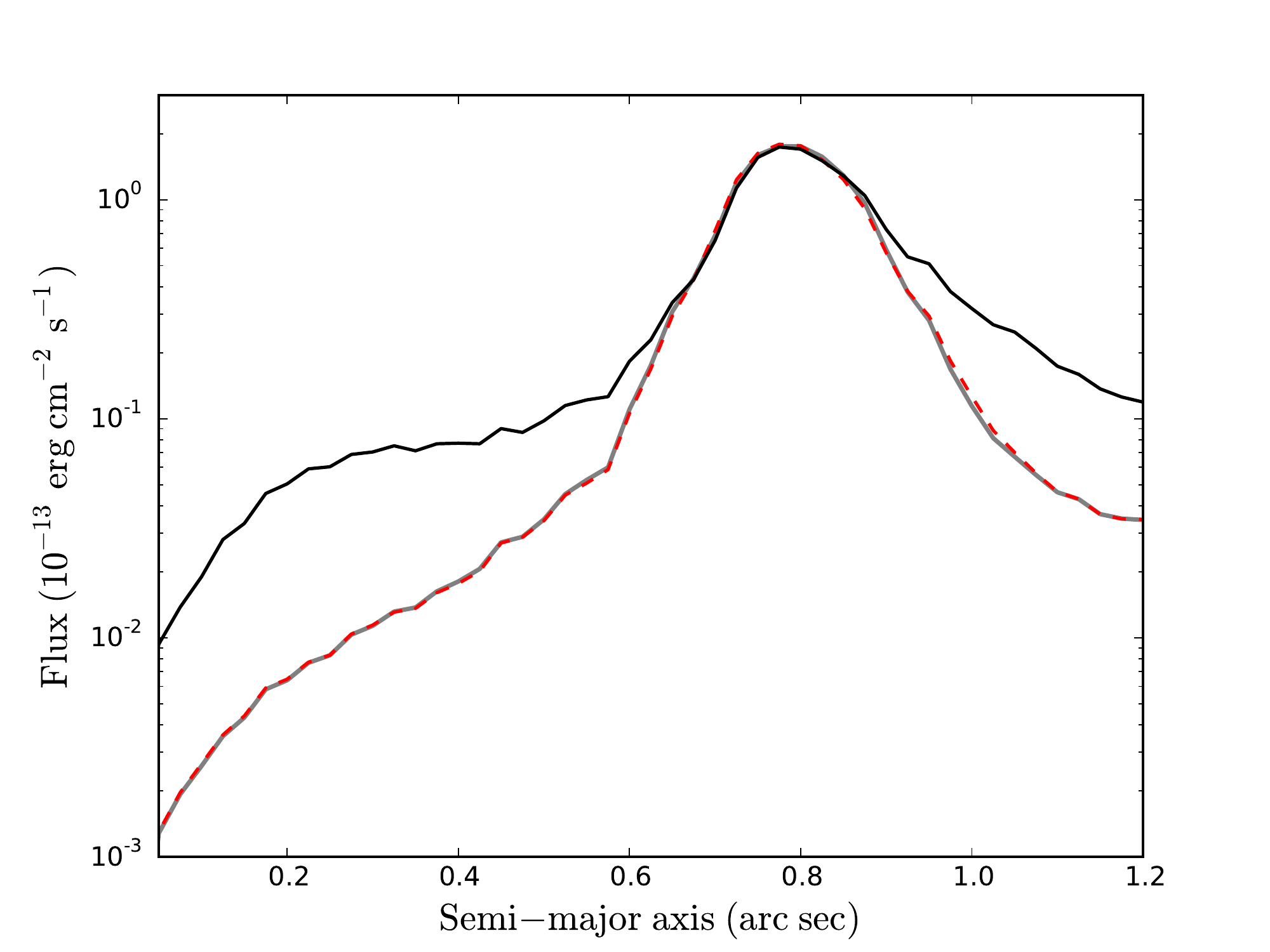}}
\caption{Flux of SN~1987A calculated in elliptical annuli for the R-band observation at 11,500 days (2018) and the corresponding ER models. The semi-major axis of the elliptical aperture was increased in steps of $0\farcs{025}$ (1 pixel).  The black line shows the observations, the grey line the model for the ER (right panel of Fig.~\ref{fig:tinymodel}), and the dashed red line the same model, but with every other point source removed (see text for details).  \\ 
\label{fig:tinymodelrad}}
\end{center}
\end{figure}

The ER in SN~1987A is more than an order of magnitude brighter than the ejecta. This means that scattered/spread light from the ER (i.e the tails of the PSFs) constitute a significant background in the ejecta region. To estimate how much this affects the flux measurements presented in Section~\ref{sec:lightcurves}, we construct synthetic models for the ER using  the {\it HST} PSF modeling tool Tiny Tim.\footnote{\url{http://www.stsci.edu/hst/observatory/focus/TinyTim}} In the modeling we approximate the ER as a collection of point sources. We choose seven epochs between 6000 and 11,500 days (listed in Table~\ref{tab:ringmodels}) and construct models in both the R and B bands. The scattered light from the ER before 6000~days is negligible. 

As a starting point for the models we simulated point sources using the positions and fluxes determined for the hotspots in Section \ref{sec:expansion}. We also convolved the model  with a Gaussian with $\sigma = 0\farcs{024}$ since we found that the point sources simulated with Tiny Tim were somewhat narrower than stars in the field. This is likely  due to the perfect sampling of the simulated PSFs, which are always positioned at the centers of pixels. Inspection of the differences between the observations and these initial models revealed clear residuals, in particular around the star superposed on the south-western part of the ER (and excluded from the measurements in Section \ref{sec:expansion}) and in the western part of the ER at late time (where some new spots appear, as seen in Figure~\ref{fig:rdiff}). We therefore added additional point sources to the models corresponding to these residuals. Finally, we normalized the models to the same total flux as the observations in a $0\farcs{2}$ wide elliptical annulus covering the ER. This correction is a the level of a few per cent. 

As an example, Figure~\ref{fig:tinymodel} shows the resulting model for the R band at 11,500~days (2018) together with the observation. The radial flux profiles of the model and observation are plotted in Figure~\ref{fig:tinymodelrad}, showing that the agreement is very good  around the peak of the ER emission. The discrepancy inside the peak is due to the ejecta and RS, while the higher observed flux outside the ER is due to a significant diffuse component, which is not included in the model. On a more detailed level there are discrepancies between the model and the observations even around the peak. Considering all fourteen models, the distribution of pixel values obtained when subtracting the model from the data has a typical standard deviation corresponding to $\sim 20 \%$ of the mean observed value for the ER. However, these differences do not significantly affect the estimates of the scattered light in the ejecta region. As a test of this, we removed every second point source from the model in Figure~\ref{fig:tinymodel}, but still normalized it to the same total flux. This results in a very bad model with alternating positive residuals and strongly oversubtracted point sources. However, the radial profile is almost identical to the original model (see Figure~\ref{fig:tinymodelrad}) and the predicted flux in the ejecta region differs by only 2~\%. The estimates of the amount of scattered light are thus not sensitive to the detailed model of the ER, as long as point sources with the correct total flux are placed in an ellipse. In Table~\ref{tab:ringmodels} we list the fraction of the flux in the ejecta region that is estimated to be due to scattered light from the ER in these models. We note that the contribution from scattered light depends both on the brightness of the ER and the size of the aperture, which expands with time (see Section~\ref{sec:lightcurves}).

\begin{deluxetable}{ccccccc}[t]
\tablecaption{Estimates of scattered light from the ER \label{tab:ringmodels}}
\tablecolumns{4}
\tablenum{7}
\tablewidth{0pt}
\tablehead{
\colhead{Date} & 
\colhead{Epoch\tablenotemark{a}} &
\colhead{B band\tablenotemark{b}} &
\colhead{R band\tablenotemark{b}} \\
\colhead{(YYYY-mm-dd)} &
\colhead{(d)} &
\colhead{(\%)} &
\colhead{(\%)} 
}
\startdata
2003-11-28 & 6122 & 3  &   13 \\
2006-12-06 & 7226 & 7 &   17 \\
2009-12-12 & 8328 &  9 &    18 \\
2011-01-05 & 8717 &  11 &    17 \\
2014-06-15 & 10,317 & 12 &   17 \\
2016-06-08 & 10,698 & 13 &    17 \\
2018-07-08 & 11,458 & 13 &   17 \\
\enddata
\tablenotetext{a}{Days since explosion on 1987-02-23.}
\tablenotetext{b}{Fraction of flux measured in the ejecta aperture that is due to scattered light from the ER according to the model.}
\end{deluxetable}

\section{VLT/UVES observations}

\label{app:uves}
\begin{figure}[t]
\begin{center}
\resizebox{70mm}{!}{\includegraphics{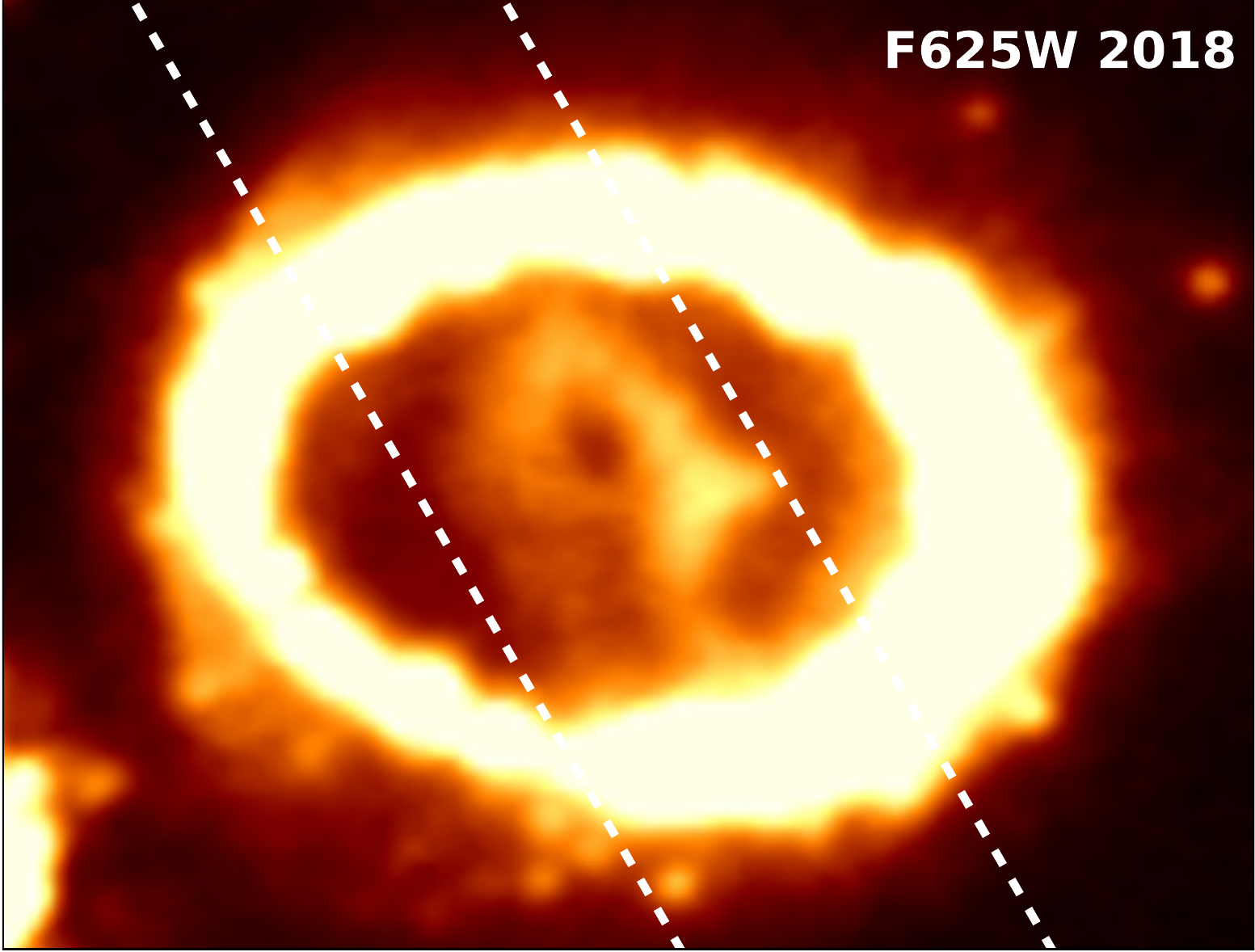}}
\caption{Slit position for the VLT/UVES observation at 11,200~days (2017) shown superposed on the WFC3/F625W image from 11,500 days (2018). The image has been scaled by an asinh function in order to highlight the weak emission outside the ER. The FOV is $2\farcs{7} \times 2\farcs{1}$. \\ 
\label{fig:slituves}}
\end{center}
\end{figure}

In this Appendix we briefly describe the VLT/UVES observations of the RS in SN~1987A, which are discussed in Section~\ref{sec:disc-rs}. The data reduction is similar to that in \cite{Fransson2013} and full details will be provided in K. Migotto at al.~(in preparation). The observations were obtained around 11,200~days (2017 Oct-Nov) with a total exposure time of 12,000~s. A $0\farcs{8}$ wide slit with a position angle of 30\degr\ was used (see Figure~\ref{fig:slituves}), which covers approximately half of the remnant.  Because the spatial resolution along the slit is limited by the seeing ($\sim 0\farcs{5} -0\farcs{6}$), we cannot obtain accurate spatial information along the slit. However, the resolution is sufficient for  extracting spectra in the northern and southern halves (shown in Figure~\ref{fig:rev_shock}). In order to determine the extension of the faint wings of the RS line profile, we subtracted a continuum which we define from a linear interpolation between the line-free regions between $\pm (17,500 - 20,000)$ \kms.



\end{document}